# Fluid Dynamics and Passive Scalar Transport Driven by Non-Uniform Tumbling of a Prolate Spheroid in Simple Shear Flow

Yanxing Wang[a,1], Hui Wan[b], Tie Wei[c], and Fangjun Shu[a]

[a]Department of Mechanical and Aerospace Engineering, New Mexico State University, Las Cruces, NM 88003;
[b]Department of Mechanical and Aerospace Engineering, University of Colorado, Colorado Springs, CO 80918;
[c]Department of Mechanical Engineering, New Mexico Institute of Mining and Technology, Socorro, NM 87801

Using high-fidelity numerical simulations based on a lattice Boltzmann framework, the advection-enhanced transport of a passive scalar from a prolate spheroid in simple shear flow has been thoroughly investigated across various parameters, including the spheroid's aspect ratio, particle-to-fluid density ratio, Reynolds number, and Schmidt number. The Reynolds number is constrained to the range of $0 \leq Re \leq 1$, where the prolate spheroid tumbles around its minor axis, aligned with the vorticity axis, in an equilibrium state. Several key findings have emerged: 1) Particle inertia significantly influences the uniformity of the spheroid's tumbling, affecting flow patterns around the spheroid and, consequently, the modes of scalar transport; 2) Both uniform and non-uniform tumbling generate a scalar line in the fluid with elevated scalar concentration, which sweeps through the wake region and merges with clusters of previously formed scalar lines; 3) Fluid passing over the spheroid carries the passive scalar downstream along these scalar lines; 4) Variations in the uniformity of spheroid tumbling result in distinct flow patterns and scalar transport modes, leading to different transport rates; 5) Within the studied parameter ranges, increased particle inertia enhances the scalar transport rate; 6) When both fluid and particle inertia are minimal, the dimensionless scalar transport rate for different aspect ratios converges to a common dependence on the Peclet number. These phenomena are analyzed in detail.

## 1. Introduction

Heat and mass transport from solid particles suspended in shear flows is fundamentally important in both scientific and industrial contexts, ranging from traditional drug delivery (Salehi et al. 2020) and metal ore heap leaching (Petersen 2016) to emerging technologies like renewable biomass energy (Badgujar & Bhanage 2015) and dissolvable microrobots (Chamolly & Lauga 2019). At low Prandtl (Pr) and Schmidt (Sc) numbers—ratios of fluid kinematic viscosity to heat and mass diffusivity—diffusion dominates the transport process. Numerous studies have explored diffusion-based transport processes, leading to the development of many analytical and theoretical models that are widely applied across various fields (Costa & Lobo 2001, Wang et al. 2012, 2022). However, in applications such as drug particle dissolution, the Schmidt number for drug molecule diffusion in water can reach as high as $O(10^4)$ (Wang et al. 2015). In these scenarios, diffusive transport becomes inefficient, and local hydrodynamics, arising from interactions between the carrier fluid and dispersed particles, provide an advective mechanism for heat and mass transport. These hydrodynamics consist of two main components: advection due to the relative velocity between particles and the surrounding fluid, and flow shear from the relative motion of adjacent flow layers. Wang and Brasseur (2019) emphasized the critical role of flow shear in mass transport for small particles, where particle rotation induces a local recirculating flow that aids in advecting dissolved mass away from the particle surface.

In nondimensional form, heat and mass transfer rates are represented by the Nusselt ($Nu$) and Sherwood numbers ($Sh$), respectively. These numbers quantify the ratio of advective to diffusive transfer rates. When the temperature or species concentration on a particle's surface is fixed, the Nusselt or Sherwood number is determined by the Reynolds ($Re$) and Peclet ($Pe$) numbers, where $Pe$ is defined as $RePr$ or $ReSc$, based on particle size and flow shear rate. Since the 1960s, significant efforts have been made to characterize heat and mass transfer from spherical particles and to establish the dependence of $Nu$ or $Sh$ on $Re$ and $Pe$. For instance, Frankel and Acrivos (1968) derived an asymptotic formula for the Sherwood number in terms of the Peclet number in the limit of $Pe$ approaching zero at $Re = 0$, expressed as $Sh = 1 + 0.257Pe^{1/2}$. For large Peclet numbers, Acrivos (1971) utilized the analogy between spheres and cylinders in simple shear flow to solve the transfer problem approximately. He discovered that at $Re = 0$, a freely suspended sphere is enveloped by a region of closed streamlines, across which heat and mass transfer occurs solely by diffusion, causing $Nu$ or $Sh$ to asymptotically approach a constant (approximately 4.5) as $Pe$ becomes infinite. Batchelor (1979) conducted a comprehensive analysis of heat and mass transfer rates from a suspended particle in a steady flow with a linear velocity distribution, deriving the same asymptotic equation as Frankel and Acrivos (1968) but with a different proportionality constant. These theoretical studies laid the groundwork for subsequent research. Using an advanced method of asymptotic interpolation on a prescribed form of the formulas for $Sh$ and $Pe$, Polyanin and Dil'man (1985) developed an approximate formula for the Sherwood number across the entire range of Peclet numbers at $Re = 0$. Subramanian and Koch (2006a, b) demonstrated that flow inertia, resulting from non-zero Reynolds numbers, disrupts the closed streamlines around a particle in simple shear flow, introducing an additional heat and mass transfer mechanism. They derived a correlation for $Sh$ as a function of $Re$ and $Pe$, applicable in the asymptotic limits of $Re$ much less than 1 and $Re$ much greater than $1/Pe^{2/5}$. However, due to the inherent complexity of transport problems, purely theoretical methods cannot fully address the effects of Reynolds numbers, and thus cannot accurately predict Nusselt and Sherwood numbers at finite Reynolds numbers.

With the rapid advancement of computational fluid dynamics methods in recent decades, large-scale numerical simulations have enabled in-depth studies of heat and mass transport around moving particles across a wide range of flow conditions. However, research on the enhancement of heat and mass transport from spherical particles driven by flow shear remains limited. A significant contribution in this area was made by Wang and Brasseur (2019), who developed a numerical model based on the lattice Boltzmann method. They conducted an extensive analysis of the hydrodynamic enhancement of heat and mass release from a spherical particle suspended in a simple shear flow, with Reynolds numbers up to 10 and Schmidt numbers up to 100. From their numerical results, they developed an accurate correlation for the shear enhancement of the Nusselt or Sherwood number as a function of Reynolds and Peclet numbers.

In realistic scenarios, it is widely recognized that most particles in nature and various applications are not perfectly spherical. The morphology of these particles significantly influences heat and mass transfer at the particle scale (Zhong et al. 2016a,b). Non-spherical particles, in particular, exhibit unique characteristics and behaviors, such as enhanced margination in shear flow, which makes them highly valuable in applications like bioparticle separation (Masaeli et al. 2012) and drug delivery (Decuzzi et al. 2009). To better understand the behavior of non-spherical particles, the rotational dynamics of a single spheroidal particle—whether oblate or prolate—freely suspended in shear flow have been the subject of extensive investigation since the 1920s, leading to several important advancements.

In 1922, Jeffery conducted pioneering work by theoretically analyzing the rotation of spheroidal particles in a simple shear flow, neglecting both fluid and particle inertia (Jeffrey 1922). He discovered that spheroids follow a family of closed paths, now known as Jeffery orbits, around the vorticity axis, with the specific orbit determined by the particle's initial orientation. Since then, numerous experimental (Taylor 1923; Binder 1939; Karnis et al. 1963; Stover & Cohen 1990) and theoretical studies (Saffman 1956; Subramanian & Koch 2005, 2006) have explored the effects of fluid and particle inertia on these orbits. Fluid inertia is characterized by the Reynolds number (Re), based on flow shear rate and particle size, while particle inertia is characterized by the Stokes number (St), defined as the product of the Reynolds number and the particle-to-fluid density ratio ($\varepsilon$). These studies have shown that fluid and particle inertia cause particles to drift towards several stable orbits, depending on the spheroid's aspect ratio. Specifically, prolate spheroids tend to drift towards a tumbling mode, rotating around their minor axes aligned with the vorticity axis, while oblate spheroids drift towards a log-rolling mode, also rotating around their minor axes aligned with the vorticity axis.

With the development of the lattice Boltzmann method (LBM) around 2000, extensive LBM-based numerical studies on the rotation dynamics of spheroidal particles have been conducted (Aidun, Lu & Ding 1998; Ding & Aidun 2000; Zettner & Yoda 2001; Subramanian & Koch 2005; Huang, Wu & Lu 2012a; Huang et al. 2012b; Mao & Alexeev 2014; Rosén, Lundell & Aidun 2014; Rosén et al. 2015a, b). These studies have identified more stable states of particle rotation at higher Reynolds numbers, including kayaking, inclined rolling, inclined kayaking, and steady states (Qi & Luo 2003; Yu, Phan-Thien & Tanner 2007; Huang, Yang, Krafczyk & Lu 2012; Rosén et al. 2015a, b). Additionally, for very heavy particles ($St \gg 1$), strong particle inertia has been found to induce a transition from a tumbling mode, where angular velocity depends on orientation, to a rotating mode with constant angular velocity (Lundell & Carlsson 2010; Nilsen & Andersson 2013). The occurrence of these states is influenced by a complex nonlinear relationship involving multiple factors, including particle aspect ratio, Reynolds number, and Stokes number. When examining the transition between various rotation modes and the drift of particles from arbitrary initial orientations to equilibrium states, additional modes of particle rotation are likely to emerge. These complex rotation modes can be accompanied by intricate fluid flows around the particles, leading to sophisticated advective transport of heat and mass between the particles and the surrounding fluid, as well as within the fluid itself. The importance of understanding these advective transport mechanisms is self-evident. However, the multitude of influencing factors creates a vast parameter space, making it a formidable challenge to fully uncover the underlying physics. Fortunately, in multiphase flows containing particles, the Reynolds number based on particle size generally remains below 10. Within this range, the equilibrium state of particle motion is restricted to tumbling for prolate spheroids and log-rolling for oblate spheroids, which significantly reduces the complexity of the scenarios that need to be considered.

Wang et al. (2023) employed a high-fidelity numerical model based on the lattice Boltzmann method to explore the advective transport of a passive scalar from an oblate spheroid undergoing log-rolling motion in a simple shear flow. Due to the axial symmetry of the particle's geometry and its steady log-rolling relative to the vorticity axis, the flow field and scalar transport around the oblate spheroid resemble those around a sphere. In contrast, a prolate spheroid exhibits periodic tumbling in its equilibrium state, leading to both geometric asymmetry with respect to the vorticity axis and non-uniform angular velocity. This non-uniform tumbling involves continuous acceleration and deceleration in rotation, resulting in a dynamic motion relative to the surrounding fluid. Additionally, this non-uniform tumbling can disrupt the spiral flow patterns typically formed on the lateral sides of uniformly rotating particles, such as spheres and oblate spheroids. Consequently, the tumbling of a prolate spheroid may result in a fundamentally different transport mechanism compared to an oblate spheroid. However, research on this transport mechanism, especially at high Peclet numbers, remains sparse in the current literature.

To enhance our understanding of heat and mass transport driven by microscale fluid and particle dynamics, and to lay the groundwork for broader research on transport mechanisms in multiphase flows, we embark on an investigation into the shear flow-driven transport mechanisms of a single prolate spheroid. In this study, we develop a high-fidelity numerical model utilizing the lattice Boltzmann framework. This model allows us to thoroughly examine the transport process of a passive scalar released from a single prolate spheroid suspended in a simple shear flow, within the Reynolds number and Stokes number ranges of $0 \leq Re \leq 1$ and $0 \leq St \leq 40$, respectively, which encompass the majority of current applications involving multiphase flows. To elucidate the advective transport mechanism, we explore a Schmidt number range from $Sc = 10$ to $100$, exceeding the typical $Sc \approx 1$ found in diffusion-dominated transport processes. We meticulously analyze the effects of fluid inertia, particle inertia, the aspect ratio of the prolate spheroid, and the diffusivity of heat and mass transport on both fluid and particle dynamics, as well as on transport efficiency. The paper is structured as follows: Section 2 introduces the physical model and defines the relevant parameters. Section 3 provides a detailed description of the numerical model based on the lattice Boltzmann method. In Section 4, we discuss the numerical results, highlighting the impact of particle inertia on the uniformity of spheroid tumbling. We characterize the flow patterns and scalar transport modes around the spheroid under both non-uniform and uniform tumbling conditions. Additionally, we quantitatively analyze the influence of the Reynolds number, particle-to-fluid density ratio, aspect ratio of the spheroid, and Peclet number on the scalar transport rate. Finally, Section 5 presents the conclusions.

## 2. Physical model

The surface of a spheroidal particle is described by

$$\frac{x'^2}{r_a^2} + \frac{y'^2}{r_b^2} + \frac{z'^2}{r_c^2} = 1 \qquad (2.1)$$

where $r_a$, $r_b$, and $r_c$ denote the lengths of three semi-principal axes, and $(x', y', z')$ are the surface coordinates of the spheroid in the body-fixed system. For a prolate spheroid, $r_a = r_b < r_c$, that is, $r_a$ and $r_b$ are defined as the length of semi-minor axis and $r_c$ is the length of semi-major axis. In addition, we define an equivalent spherical radius of the prolate spheroid using the radius of a sphere of the same volume,

$$R \equiv (r_a r_b r_c)^{1/3} \qquad (2.2)$$

The aspect ratio is defined as,

$$\Lambda \equiv \frac{r_c}{r_a} \qquad (2.3)$$

We focus on the transport process in the equilibrium state of particle rotation in which the prolate spheroid tumbles with a periodic speed about its minor axis, which is aligned with the vorticity vector of the background flow. Gravity and buoyancy are ignored. Figure 1 shows the physical model of numerical study, in which a single prolate spheroid is placed symmetrically on the central plane of an incompressible flow confined by two parallel planes. The two planes move in opposite directions at the same speed $U_0$ to produce flow shear. The distance between the two plates is $2H$, and the length and width of the planes are $2L$ and $2W$, respectively. The background flow shear rate, denoted by $G$, is calculated from $U_0$ and $H$ as:

$$G = \frac{U_0}{H} \tag{2.4}$$

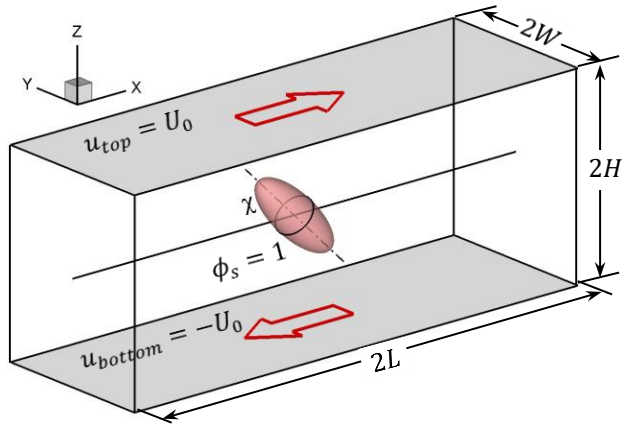

FIGURE. 1. Physical model of a prolate spheroid neutrally suspended in a simple shear flow with passive scalar released at spheroid surface. The simple shear flow is created by two parallel planes moving in opposite directions. The horizontal velocities of top and bottom planes are fixed at $u_{top} = U_0$ and $u_{bottom} = -U_0$, respectively. The concentration of passive scalar is fixed at $\phi_s = 1$ at the surface of prolate spheroid.

The particle is allowed to move in response to surface pressure and shear stress exerted by the surrounding fluid. In several important studies, the Reynolds number is defined using the length of semi-major axis as the characteristic length (Ding & Aidun 2000; Qi & Luo 2003; Huang et al. 2012; Mao & Alexeev 2014; Rosén et al. 2015b). In this work, we examine the impact of particle aspect ratio on flow and transport modes in the same flow environment, it is more convenient to define a unified Reynolds number for particles of the same volume. Therefore, we define the Reynolds number based on the equivalent spherical radius $R$ (Eqn. 2.2) and the background flow shear rate $G$ (equation (2.4)),

$$Re \equiv \frac{GR^2}{\nu} \tag{2.5}$$

where $\nu$ is the kinematic viscosity of the carrier fluid. When considering the effect of particle inertia, particle-to-fluid density ratio $\varepsilon$ is an important nondimensional parameter, which is defined as,

$$\varepsilon \equiv \frac{\rho_p}{\rho_f} \tag{2.6}$$

where $\rho_p$ is the density of spheroid and $\rho_f$ the density of carrier fluid. Particle inertia is characterized by the Stokes number $St$, which is defined as the product of particle-to-fluid density ratio $\varepsilon$ and Reynolds number $Re$,

$$St \equiv \varepsilon Re \tag{2.7}$$

Since our definition of Reynolds number (equation (2.5)) is different from that in previous literature, our Stokes number is also different. In the analysis we use particle-to-fluid density ratio rather than Stokes number.

To predict heat and mass transport, temperature and mass concentration is modeled as the concentration $\phi$ of a passive scalar released at particle surface, where $\phi$ is fixed at $\phi_s = 1$. Initially $\phi = 0$ everywhere in the flow. The Peclet number, $Pe$, which characterizes the effect of flow shear on scalar transport, is defined as,

$$Pe \equiv \frac{GR^2}{D} = ReSc \tag{2.8}$$

where $D$ is the diffusivity of scalar transport and $Sc = \nu/D$ is the Schmidt number of passive scalar in the fluid. To compare the release rate of passive scalar from particles with different aspect ratios, we define the nondimensional release rate, that is, the Sherwood number ($Sh$) based on the radius and surface area of a sphere with the same volume as the prolate spheroid,

$$Sh \equiv \frac{QR}{DA\phi_s} \tag{2.9}$$

where $Q$ is the release rate of passive scalar from spheroid surface, $A$ ($= 4\pi R^2$) the surface area of the sphere, and $\phi_s$ ($= 1$) the concentration of passive scalar at spheroid surface. This definition highlights the change in release efficiency resulting from variations in particle shape, while the definition based on the actual surface area of particles in traditional diffusion transport studies emphasizes the release rate per unit area.

A complete description of the problem includes the Reynolds number $Re$, Schmidt number $Sc$, particle-to-fluid density ratio $\varepsilon$ or Stokes number $St$, aspect ratio of prolate spheroid $\Lambda$, and the ratios of the height, width, and length of the computational domain to the equivalent spherical radius ($2H/R$, $2W/R$, and $2L/R$). We align our study with common natural and technological scenarios by setting the ratio of the semi-major axis length to the equivalent spherical radius, denoted as $r_c/R$, within the range of $1 \leq r_c/R \leq 4$, which corresponds to aspect ratios of $1 \leq \Lambda \leq 8$. While Wang et al. (2022) have extensively addressed diffusion-dominated transport of passive scalars from a prolate spheroid, our study focuses on the advective enhancement of scalar transport. Reflecting realistic applications such as drug dissolution, we select a Reynolds number range from $O(10^{-2})$ to $O(1)$, where the influence of fluid inertia transitions from weak to comparable with fluid viscosity. We also examine a particle-to-fluid density ratio range from $O(10^{-1})$ to $O(10)$, covering the spectrum from weak to strong particle inertia, with corresponding Stokes numbers ranging from $O(10^{-3})$ to $O(10)$. Due to the high grid resolution required for simulating transport processes with large $Sc$, the Schmidt number is confined to the range of $O(10^1)$ to $O(10^2)$. Periodic boundary conditions for flow velocity and scalar concentration are applied at the streamwise and spanwise boundaries. To minimize the impact of computational boundaries, the computational domain's height, width, and length are set to be over an order of magnitude larger than the equivalent sphere radius, with ratios of $2H/R = 20$, $2W/R = 20$, and $2L/R = 140$. These ratios exceed those used in previous studies (Ding & Aidun 2000; Qi & Luo 2003; Huang et al. 2012; Mao & Alexeev 2014; Rosén et al. 2015b). Table 1 provides a summary of the parameters considered in this study.

Table 1. Case parameters considered in this study.

| $r_c/R$ | $\Lambda\ (= r_c/r_a)$ | $\varepsilon\ (= \rho_p/\rho_f)$ | $Re$ | $Sc$ | $2H/R$ | $2W/R$ | $2L/R$ |
|---|---|---|---|---|---|---|---|
| 1, 2, 3, 4 | 1, 2.83, 5.20, 8 | 0.4, 4, 8, 12, 16, 20, 30, 40 | 0.05, 0.1, 0.3, 0.5, 0.75, 1 | 10, 20, 40, 60, 80, 100 | 20 | 20 | 140 |

At the start of each simulation, the major axis of the prolate spheroid is aligned on the shear plane and oriented perpendicular to the flow direction ($\chi_0 = \pi/2$). The spheroid then moves freely, propelled by the surrounding fluid. Following an initial adjustment period, the motion of both the fluid and the spheroid, along with the scalar transport, reaches an equilibrium state. In this state, the prolate spheroid tumbles on the shear plane around the vorticity axis with a time-periodic angular velocity. Given that the Schmidt number exceeds 1 in all scenarios examined in this study, the flow velocity stabilizes more quickly than the concentration of the passive scalar. Analysis is conducted once the time-averaged scalar release rate stabilizes to a constant value. Despite the particle's translational motion being unrestricted, the major axis consistently remains on the central shear plane, and the particle's center stays at its initial position.

## 3. Numerical methods

A 3D high-fidelity numerical model based on a lattice Boltzmann method is utilized in the present study because it is highly parallelizable and highly capable in dealing with stationary and moving solid boundaries with complex geometries. The dependent variable is the particle distribution function $\mathbf{f}(\mathbf{x}, t)$, which quantifies the probability of finding an ensemble of molecules at position $\mathbf{x}$ with velocity $\mathbf{e}$ at time $t$ (Qian 1992; Chen & Doolen 1998; Wang et al. 2010). Continuum-level velocity $\mathbf{u}(\mathbf{x}, t)$ and density $\rho(\mathbf{x}, t)$ are obtained from moments of $\mathbf{f}(\mathbf{x}, t)$ over velocity space. In three dimensions, the velocity vector $\mathbf{e}$ can be discretized into 15, 19, or 27 components (referred to as D3Q15, D3Q19 and D3Q27) (Qian & d'Humieres 1992). Here we apply the D3Q19 approach, mainly to reduce computational load, with the recognition that the Reynolds number is relatively low.

The LB equation with the Bhatnagar-Gross-Krook (Bhatnagar et al. 1954) representation for the collision operator is (Chen & Doolen 1998) is written as,

$$\mathbf{f}(\mathbf{x} + \mathbf{e}\delta_t, t + \delta_t) - \mathbf{f}(\mathbf{x}, t) = -\frac{1}{\tau}\big(\mathbf{f}(\mathbf{x}, t) - \mathbf{f}^{eq}(\mathbf{x}, t)\big) \tag{3.1}$$

where the discretized velocity vector **e** for D3Q19 is,

$$\mathbf{e}_\alpha = \begin{cases} (0,0,0) & \alpha = 0 \\ (\pm 1,0,0), (0,\pm 1,0), (0,0,\pm 1) & \alpha = 1,2,\cdots,6 \\ (\pm 1,\pm 1,0), (\pm 1,0,\pm 1), (0,\pm 1,\pm 1) & \alpha = 7,8,\cdots,18 \end{cases} \tag{3.2}$$

where α is the index of components of velocity vector **e**. The left-hand side (LHS) of equation (3.1) describes "streaming," the exchange of momentum between neighboring lattice nodes as a result of bulk advection and molecular diffusion. The right-hand side (RHS) describes the mixing, or collision of molecules that drive the distribution function ($\mathbf{f}(\mathbf{x}, t)$) toward the equilibrium distribution function ($\mathbf{f}^{eq}(\mathbf{x}, t)$) with a relaxation time scale, τ. The relaxation time is linearly related to the fluid kinematic viscosity ν by

$$\nu = (2\tau - 1)c\delta x/6 \tag{3.3}$$

where $\delta x$ is the lattice spacing and $c = \delta x/\delta t$ is the basic speed on the lattice. In the low Mach number limit, the equilibrium distribution for the D3Q19 model is (Qian & d'Humieres 1992)

$$\mathbf{f}^{eq}(\mathbf{x}, t) = w_\alpha \rho(\mathbf{x}, t)\left[1 + 3\frac{\mathbf{e}_\alpha \cdot \mathbf{u}}{c^2} + \frac{9}{2}\frac{(\mathbf{e}_\alpha \cdot \mathbf{u})^2}{c^4} - \frac{3}{2}\frac{(\mathbf{u} \cdot \mathbf{u})^2}{c^2}\right] \tag{3.4}$$

where $w_\alpha$ are weighting factors, $w_0 = 1/3$, $w_\alpha = 1/18$ for $\alpha = 1 - 6$ and $w_\alpha = 1/36$ for $\alpha = 7 - 18$. The continuum-level fluid density $\rho(\mathbf{x}, t)$ and momentum $\rho\mathbf{u}(\mathbf{x}, t)$ are obtained from the discretized moments of the particle distribution function,

$$\rho(\mathbf{x}, t) = \sum_\alpha f_\alpha(\mathbf{x}, t), \quad \rho(\mathbf{x}, t)\mathbf{u}(\mathbf{x}, t) = \sum_\alpha f_\alpha(\mathbf{x}, t)\mathbf{e}_\alpha \tag{3.5}$$

In the treatment of moving particle surfaces, we use 2nd order accurate scheme advanced by Lallemand and Luo (2003). This method is based on the simple bounce-back boundary scheme and interpolations. If the distance fraction of the first fluid node from the solid boundary, *q*, is less than half the lattice space, the computational quantities are interpolated before propagation and bounce-back collision. If *q* is greater than half lattice space, interpolation is conducted after propagation and bounce-back collision. The momentum exerted by the moving boundary is determined by these terms, which were established by Ladd (1994) and Bouzidi et al. (2001).

The motion of the spheroid is obtained by solving Newton's equations of motion,

$$M \, d\mathbf{U}(t)/dt = \mathbf{F}(t) \tag{3.6}$$

$$\mathbf{I} \cdot d\mathbf{\Omega}(t)/dt + \mathbf{\Omega}(t) \times [\mathbf{I} \cdot \mathbf{\Omega}(t)] = \mathbf{T}(t) \tag{3.7}$$

where *M* is the mass of the solid spheroid, **I** the moment of inertial tensor, **U** the translational velocity, **Ω** is the angular velocity, and **F** and **T** are the total force and torque on the particle, respectively. At each fluid boundary node $\mathbf{x}_b$, the force is calculated by the exchange of distribution function between the fluid and solid.

$$\mathbf{F}_b(\mathbf{x}_b, t) = -\sum_\alpha \big[f_{\bar{\alpha}}(\mathbf{x}_b, t + \delta_t) + \hat{f}_\alpha(\mathbf{x}_b, t_+)\big]\mathbf{e}_\alpha \tag{3.8}$$

where $f_{\bar{\alpha}}$ is the distribution transferred from the solid boundary including the additional momentum due to boundary motion, $\hat{f}_\alpha$ is the post-collision distribution transferred from the flow to the boundary, and $t_+$ is a time after collision but before streaming. According to Aidun et al. (1998), the total force includes three components, the force due to the communication of the distribution function ($\mathbf{F}_b(\mathbf{x}_b, t)$), the force due to momentum transfer from fluid to solid when some grid nodes are covered by solid ($\mathbf{F}_c(\mathbf{x}, t)$), and the force due to momentum transfer from solid to fluid when some grid nodes are uncovered by solid ($\mathbf{F}_u(\mathbf{x}, t)$). The latter two are given as,

$$\mathbf{F}_c(\mathbf{x}_c, t) = -\sum_\alpha [f_\alpha(\mathbf{x}_c, t)\mathbf{e}_\alpha] \tag{3.9}$$

$$\mathbf{F}_u(\mathbf{x}_u, t) = -\rho(\mathbf{x}_u, t)\mathbf{u}(\mathbf{x}_u, t) \tag{3.10}$$

The total force and torque are calculated by summing the force and torque at each fluid boundary node and each covered and uncovered fluid node,

$$\mathbf{F}(t) = \sum_{FBN} \mathbf{F}_b(\mathbf{x}_b, t) + \sum_{CN} \mathbf{F}_c(\mathbf{x}_c, t) + \sum_{UN} \mathbf{F}_u(\mathbf{x}_u, t) \tag{3.11}$$

$$\mathbf{T}(t) = \sum_{FBN}(\mathbf{x}_b - \mathbf{X}_o) \times \mathbf{F}_b(\mathbf{x}_b, t) \tag{3.12}$$

where FBN, CN, and UN denote the fluid boundary nodes, covered, and uncovered nodes. $\mathbf{X}_o$ is the central coordinate of the solid particle.

To address the passive scalar transport process, the moment propagation method developed by Frenkel and Ernst (1989), Lowe and Frenkel (1995), and Merks et al. (2002) is employed. In this method, a scalar quantity is released in the lattice and a scalar concentration field variable is propagated at the continuum level for each scalar using the distribution function. Letting $\phi(\mathbf{x}, t)$ be the scalar concentration on the lattice at location $\mathbf{x}$ at time $t$, the advancement of scalar concentration at the next step is given by

$$\phi(\mathbf{x}, t + \delta t) = \sum_\alpha P_\alpha(\mathbf{x} - \mathbf{e}_\alpha \delta t, t + \delta t) + \phi(\mathbf{x}, t)\Delta^* \tag{3.13}$$

$$P_\alpha(\mathbf{x} - \mathbf{e}_\alpha \delta t, t + \delta t) = \left[\frac{\hat{f}_\alpha(\mathbf{x} - \mathbf{e}_\alpha \delta t, t_+)}{\rho(\mathbf{x} - \mathbf{e}_\alpha \delta t, t)} - w_\alpha \Delta^*\right] \phi(\mathbf{x} - \mathbf{e}_\alpha \delta t, t) \tag{3.14}$$

where $\Delta^*$ is the fraction of $\phi(\mathbf{x}, t)$ that remains on the lattice node during the time advancement. $\Delta^*$ is related to the molecular diffusivity of the passive scalar as,

$$\Delta^* = 1 - 6D/c\delta x \tag{3.15}$$

The upper bound of the Peclet number is significantly higher in the momentum propagation method than in other methods, resulting in a more stable calculation.

The fixed-scalar boundary condition is enforced on the surface of the particle as well as the top and bottom plates. Specifically, on the particle surface, $\phi$ is set to $\phi_s = 1$, while on the top and bottom plates, $\phi$ is maintained at $\phi_{top} = \phi_{bottom} = 0$. The details of the implementation of boundary conditions are given in Wang et al. (2010).

The spatial resolution of the grid has consistently posed challenges in numerical simulations of freely suspended particles. While previous studies have shown qualitative agreement, quantitative discrepancies persist (Qi & Luo 2003; Yu et al. 2007; Huang et al. 2012a; Rosén et al. 2014). For instance, the critical Reynolds numbers indicating state transitions vary across studies. It is believed that certain rotational parameters are particularly sensitive to grid resolution near the particle surface, leading to these quantitative differences (Rosén et al. 2015b). In this work, we standardize grid resolution based on the Reynolds number for particles with varying aspect ratios. Specifically, for $Re < 0.5$, we use 16 grid nodes to resolve the radius of the equivalent sphere. For a particle with an aspect ratio of 8 ($r_c/R = 4$), this translates to 8 grid nodes along the semi-minor axis and 64 along the semi-major axis. For $0.5 \leq Re \leq 1$, 20 grid nodes are employed over the equivalent sphere's radius. Notably, our grid resolution is significantly higher than that of previous studies for Reynolds numbers on the order of $O(10^2)$ (Qi & Luo 2003; Yu et al. 2007; Huang et al. 2012b; Mao & Alexeev 2014; Rosén et al. 2015b). Additionally, since our focus is on fluid flow within $Re \leq 1$, the sensitivity of results to grid resolution is reduced compared to earlier literature. We have thoroughly examined the grid sensitivity of our model, finding that changes in grid resolution have an acceptable impact on flow velocity and scalar concentration. For brevity, the grid independence study is not included here. In the subsequent analysis, we will validate our numerical results against theoretical and numerical findings from the literature.

## 4. Results and discussion

The two primary questions we aim to address are: which parameters dictate the behaviors of fluid and particles, as well as the characteristics of scalar transport within a multi-parameter space, and whether this entire parameter space can be segmented into simpler regions based on transport characteristics. We propose that even in simple periodic tumbling scenarios, variations in angular velocity patterns can result in distinct transport mechanisms. When the Peclet number ($Pe$) is less than 1, scalar transport is predominantly governed by diffusion, a phenomenon extensively explored in the literature (Wang et al. 2012, 2015, 2023). However, when $Pe$ is around 1 or higher, flow advection becomes a critical factor. The behaviors of fluid and particles are primarily influenced by three dimensionless parameters: the length of the semi-major axis $r_c/R$ (or aspect ratio $\Lambda$), the Reynolds number ($Re$), and the particle-to-fluid density ratio $\varepsilon$ (or Stokes number $St$). Analyzing the variations in angular velocity of the prolate spheroid during tumbling within the parameter space defined by $r_c/R$, $Re$, and $\varepsilon$ will yield valuable insights for identifying and classifying advective transport characteristics. The subsequent analysis will be divided into three sections: the tumbling dynamics of prolate spheroids influenced by multiple parameters, the typical advective transport modes of passive scalars induced by these dynamics, and a quantitative analysis of scalar transport under the influence of various factors.

### *4.1 Mapping of Tumbling Characteristics of Prolate Spheroid*

In this section, we seek to pinpoint the key characteristics of prolate spheroid tumbling across different parameter combinations. By doing so, we aim to segment the parameter space defined by $r_c/R$, $Re$, and $\varepsilon$ into distinct regions based on these tumbling features. This segmentation will lay the groundwork for characterizing advective transport in the subsequent section.

We begin by examining the effect of the Reynolds number, considering two particle-to-fluid density ratios: $\varepsilon = 0.4$ and $\varepsilon = 40$. As we will demonstrate in the subsequent discussion (see Figure 6), for a prolate spheroid with an aspect ratio of $r_c/R = 3$, variations in $\varepsilon$ significantly impact particle rotation primarily within the range of approximately $4 \leq \varepsilon \leq 20$. Outside this range, specifically when $\varepsilon \leq 4$ or $\varepsilon \geq 20$, changes in $\varepsilon$ have a negligible effect on particle rotation.

Figure 2 illustrates the dimensionless angular velocity $\Omega/G$ of a prolate spheroid with a length of semi-major axis $r_c/R = 3$ as functions of dimensionless time $tG$ and rotation angle $\chi/\pi$, for Reynolds numbers $Re = 0.05, 0.1, 0.5$, and 1. Here, $t$ represents the time elapsed since the spheroid began moving, $\chi$ is the angle between one semi-major axis and the streamwise axis (as shown in Figure 1), and $\Omega \equiv d\chi/dt$. The results are compared with Jeffery's theoretical predictions for $Re = 0$ (Jeffery 1922). After an initial transient phase, $\Omega/G$ exhibits periodic behavior with respect to both $tG$ and $\chi/\pi$ across all cases. At $Re = 0.05$, the numerical results closely align with Jeffery's theory for both density ratios, partially validating the numerical model developed in this study. In each scenario, the duration of small $\Omega/G$ is longer than that of larger $\Omega/G$ (Figure 2(a,b)), a finding consistent with previous studies (Ding & Aidun 2000; Qi & Luo 2003) and crucial for our subsequent analysis of scalar transport rates. For $\varepsilon = 0.4$, where particle inertia characterized by $St(= \varepsilon Re)$ is weak, increasing $Re$ from 0.05 to 1 results in only a slight decrease in the maximum and minimum angular velocities ($\Omega_{max}/G$ and $\Omega_{min}/G$), leading to an increased rotation period (Figure 2(a)). These phenomena have been observed and analyzed over a broad range of Reynolds numbers (Ding & Aidun 2000; Qi & Luo 2003; Huang et al. 2012b; Mao & Alexeev 2014), and our results qualitatively agree with theirs. For $\varepsilon = 40$, enhanced particle inertia becomes more significant than fluid inertia. As $Re$ increases, so does $St$, resulting in a stronger tendency to maintain a constant $\Omega/G$.

Consequently, $\Omega_{max}/G$ decreases and $\Omega_{min}/G$ increases noticeably as $Re$ rises from 0 to 1, leading to a reduced rotation period (Figure 2(b)). These observations are consistent with the findings of Mao & Alexeev (2014).

We introduce the concept of 'uniformity' to describe the variation in angular velocity during spheroid tumbling. The deviation of a prolate spheroid's aspect ratio from that of a sphere causes the angular velocity to shift from a constant ($\Omega_{sphere}/G \approx 0.5$) to a periodic mode, resulting in nonuniform tumbling. This nonuniformity has two aspects. The first is the amplitude of variation in $\Omega/G$, represented by $\Omega_{max}/G - \Omega_{min}/G$. This amplitude of variation can also be represented by the deviation of $\Omega_{max}/G$ and $\Omega_{min}/G$ from the constant angular velocity of the sphere, that is, by $|\Omega_{max}/G - \Omega_{sphere}/G|$ and $|\Omega_{min}/G - \Omega_{sphere}/G|$. The second aspect is the difference in time spans for larger and smaller angular velocities. In all cases studied, the time span for smaller $\Omega/G$ exceeds that for larger $\Omega/G$, with the difference related to $\Omega_{min}/G$. A smaller $\Omega_{min}/G$ results in a larger time span difference. As shown in Figure 2(a) and (b), when fluid inertia dominates ($\varepsilon = 0.4$), increased fluid inertia reduces the uniformity of spheroid tumbling. Conversely, when particle inertia dominates ($\varepsilon = 40$), increased fluid inertia enhances particle inertia, improving tumbling uniformity. These phenomena align with the effects of fluid and particle inertia summarized by Rosén et al. (2015b).

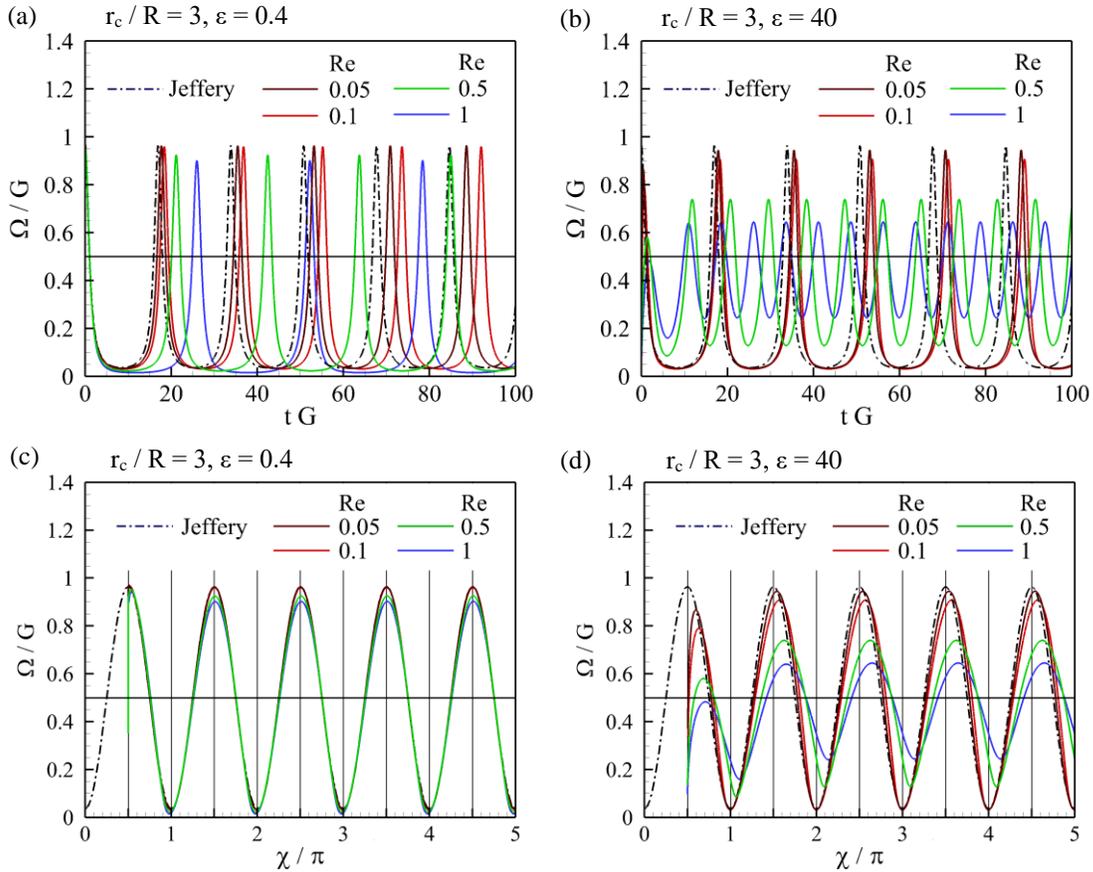

FIGURE 2. Variation of normalized angular velocity ($\Omega/G$) with normalized time ($tG$) (a and b) and rotation angle ($\chi/\pi$) (c and d) of a prolate spheroid with $r_c/R = 3$ in a simple shear for various Reynolds numbers. (a) $\Omega/G$ vs. $tG$ for $\varepsilon = 0.4$, (b) $\Omega/G$ vs. $tG$ for $\varepsilon = 40$, (c) $\Omega/G$ vs. $\chi/\pi$ for $\varepsilon = 0.4$, and (d) $\Omega/G$ vs. $\chi/\pi$ for $\varepsilon = 40$. The solid lines are the results of LBM simulation for $Re > 0$ and the dash-dot lines are the prediction of Jeffery's theory for $Re = 0$ (Jeffery 1922).

The variation of $\Omega/G$ with $\chi/\pi$, as depicted in figures 2(c) and 2(d), clearly demonstrates a dependence of $\Omega/G$ on $\chi/\pi$. This relationship does not conform to a sinusoidal pattern, even at $Re = 0$ (Jeffery 1922). When $\varepsilon = 0.4$, particle inertia is weak, and the fluid inertia within the range $0 < Re \leq 1$ is insufficient to significantly alter the symmetric dependence observed at $Re = 0$. The angular velocity $\Omega/G$ reaches its peak at $\chi/\pi \approx n + 1/2$ and its trough at $\chi/\pi \approx n$, where $n$ is an integer. This means that the angular velocity is maximized when the major axis of the prolate spheroid is perpendicular to the flow direction and minimized when it aligns with the flow direction. Conversely, when $\varepsilon = 40$, particle inertia plays a crucial role in influencing the variation of $\Omega/G$. As $Re$ increases from 0 to 1, the enhanced particle inertia induces an 'overshoot' in particle motion, causing the angles at which $\Omega_{max}/G$ and $\Omega_{min}/G$ occur to shift to larger values than $\chi/\pi \approx n + 1/2$ and $\chi/\pi \approx n$, respectively. Simultaneously, $\Omega/G$ becomes more uniform.

Figure 3 illustrates how the minimum and maximum angular velocities, $\Omega_{min}/G$ and $\Omega_{max}/G$, vary with increasing Reynolds number $Re$ for three different semi-major axes ratios, $r_c/R = 2$, 3 and 4, and two density ratios, $\varepsilon = 0.4$ and 40. The corresponding aspect ratios are $\Lambda = 2.83$, 5.20, and 8, respectively. Although the trends in $\Omega_{min}/G$ and $\Omega_{max}/G$ are similar across different $r_c/R$ values, an increase in $r_c/R$ results in a decrease in $\Omega_{min}/G$ and an increase in $\Omega_{max}/G$, indicating that higher $r_c/R$ and $\Lambda$ lead to greater nonuniformity in $\Omega/G$. The impact of particle aspect ratio will be further analyzed in Figures 7 and 8. For each $r_c/R$, when $\varepsilon = 0.4$, particle inertia is weak, and increasing $Re$ only slightly reduces both $\Omega_{max}/G$ and $\Omega_{min}/G$. Notably, $\Omega_{min}/G$ approximately follows a linear relationship with $Re$, as reported by Ding & Aidun (2000). When $\varepsilon = 40$, a critical Reynolds number $Re_c$ is identified around $Re \approx 0.1$. Below this threshold, both fluid and particle inertia are weak, resulting in minimal changes in $\Omega_{max}/G$ and $\Omega_{min}/G$. Consequently, $\Omega_{min}/G$ decreases as $Re$ increases from 0 to 0.1, as shown in Figure 3(a). In the higher Reynolds number range ($Re > Re_c$), the enhanced particle inertia with increasing $Re$ significantly raises $\Omega_{min}/G$ and lowers $\Omega_{max}/G$, causing the angular velocity to approach a more uniform value. Further increases in particle inertia may lead to a transition to another equilibrium state known as rotating (Lundell & Carlsson 2010; Nilsen & Anderson 2013; Rosén et al. 2015b), a topic beyond the scope of this paper.

Figure 4 illustrates how the dimensionless period of particle tumbling, denoted as $TG$, varies with the Reynolds number for $r_c/R = 2$, 3 and 4, and density ratios $\varepsilon = 0.4$ and 40. The curves for different $r_c/R$ display similar trends, with an increase in $r_c/R$ leading to a rise in in $TG$ at each $Re$ and $\varepsilon$. This is primarily due to the decrease in $\Omega_{min}/G$ as $Re$ increases, as depicted in Figure 3(a). For each $r_c/R$, when

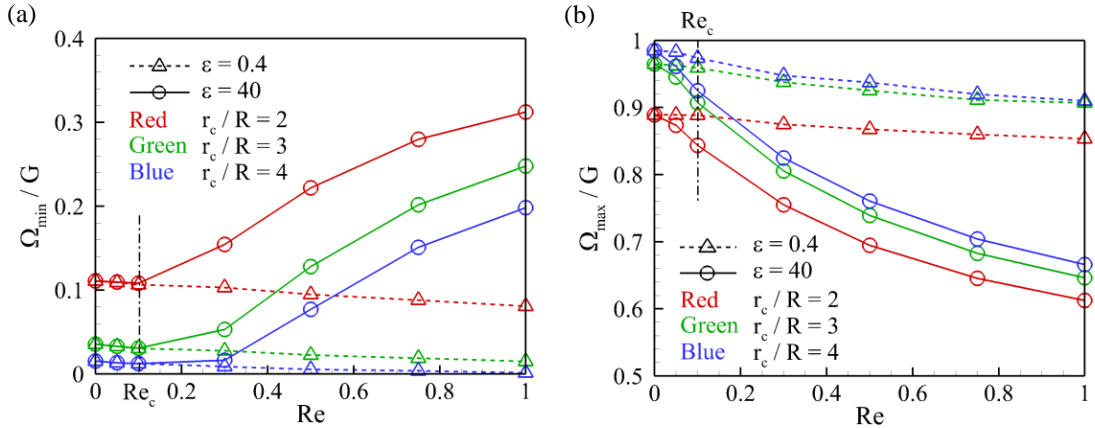

FIGURE 3. Variation of normalized minimum and maximum angular velocity ($\Omega_{min}/G$ and $\Omega_{max}/G$) with Reynolds number ($Re$) for various density ratios ($\varepsilon$) and major axes ($c/R$). (a) $\Omega_{min}/G$ vs. $Re$, (b) $\Omega_{max}/G$ vs. $Re$. Cases of $Re = 0$ are from Jeffery's theory (Jeffery 1922).

$Re < Re_c \approx 0.1$, both fluid and particle inertia are weak, resulting in $TG$ remaining largely constant over $Re$ and showing little difference between $\varepsilon = 0.4$ and 40. However, when $Re > Re_c$, the effects of fluid and particle inertia become significant. At $\varepsilon = 0.4$, where particle inertia is weak, an increase in $Re$ leads to an increase in $TG$ due to the decrease in both $\Omega_{min}/G$ and $\Omega_{max}/G$. The sensitivity of $TG$ to $Re$ grows with increasing $r_c/R$, meaning that changes in $Re$ result in larger changes in $TG$ for larger $r_c/R$. Assuming $TG$ is primarily determined by $\Omega_{min}/G$, a quadratic relationship between $TG$ and $Re$ can be derived (Ding & Aidun 2000). When $\varepsilon = 40$, where particle inertia is strong, an increase in $Re$ results in a higher $St$, making the angular velocity $\Omega/G$ more uniform. Consequently, the increase in $\Omega_{min}/G$ leads to a decrease in the rotation period $TG$, as shown in the figure.

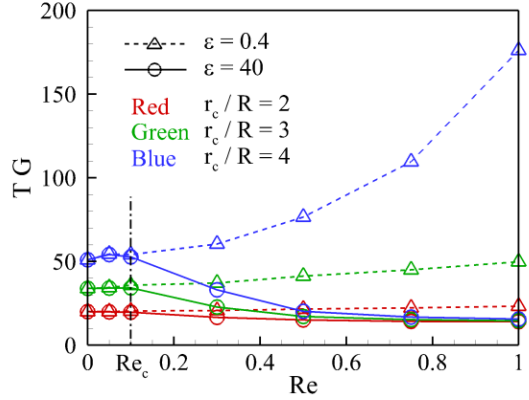

FIGURE 4. Variation of normalized rotation period ($TG$) with Reynolds number ($Re$) for various density ratios ($\varepsilon$) and major axes ($c/R$). Cases of $Re = 0$ are from Jeffery's theory (Jeffery 1922).

The analysis above demonstrates that fluid inertia and particle inertia have opposing effects on the uniformity of prolate spheroid tumbling. This aligns with the findings of Rosén et al. (2015b), which suggest that fluid inertia tends to increase the tumbling period, thereby making the motion more non-uniform, as defined in this study, and pushing the spheroid towards a steady state. Conversely, particle inertia encourages the spheroid to tumble with a constant angular velocity. Our results further suggest that fluid inertia predominantly influences spheroid tumbling only when particle inertia is weak. In essence, particle inertia plays a crucial role in determining the uniformity of the spheroid's tumbling motion.

Next, we explore the impact of the particle-to-fluid density ratio, $\varepsilon$, within the range of $0.4 \leq \varepsilon \leq 40$. Figure 5 presents the dimensionless angular velocity, $\Omega/G$, of a prolate spheroid with $r_c/R = 3$ plotted against dimensionless time, $tG$, and rotation angle, $\chi/\pi$, for $\varepsilon = 0.4, 4, 20$, and 40. At $Re = 0.1$, both fluid and particle inertia are weak, resulting in variations of $\Omega/G$ that closely align with Jeffery's theoretical predictions for $Re = 0$ (figures 5(a) and 5(c)). The particle inertia causes a slight reduction in $\Omega_{max}/G$ and a minor shift in the angle $\chi/\pi$ where $\Omega_{max}/G$ occurs. These effects become more pronounced as $\varepsilon$ increases. At $Re = 1$, the increased fluid inertia enhances particle inertia, leading to significant changes in $\Omega/G$ as depicted in figures 5(b) and 5(d). Notable changes include a decrease in $\Omega_{max}/G$, an increase in $\Omega_{min}/G$, and a reduction in the rotation period. Additionally, the angles at which $\Omega_{min}/G$ and $\Omega_{max}/G$ occur increase with higher $\varepsilon$. The underlying reasons for these changes have been discussed in the analysis of figure 2.

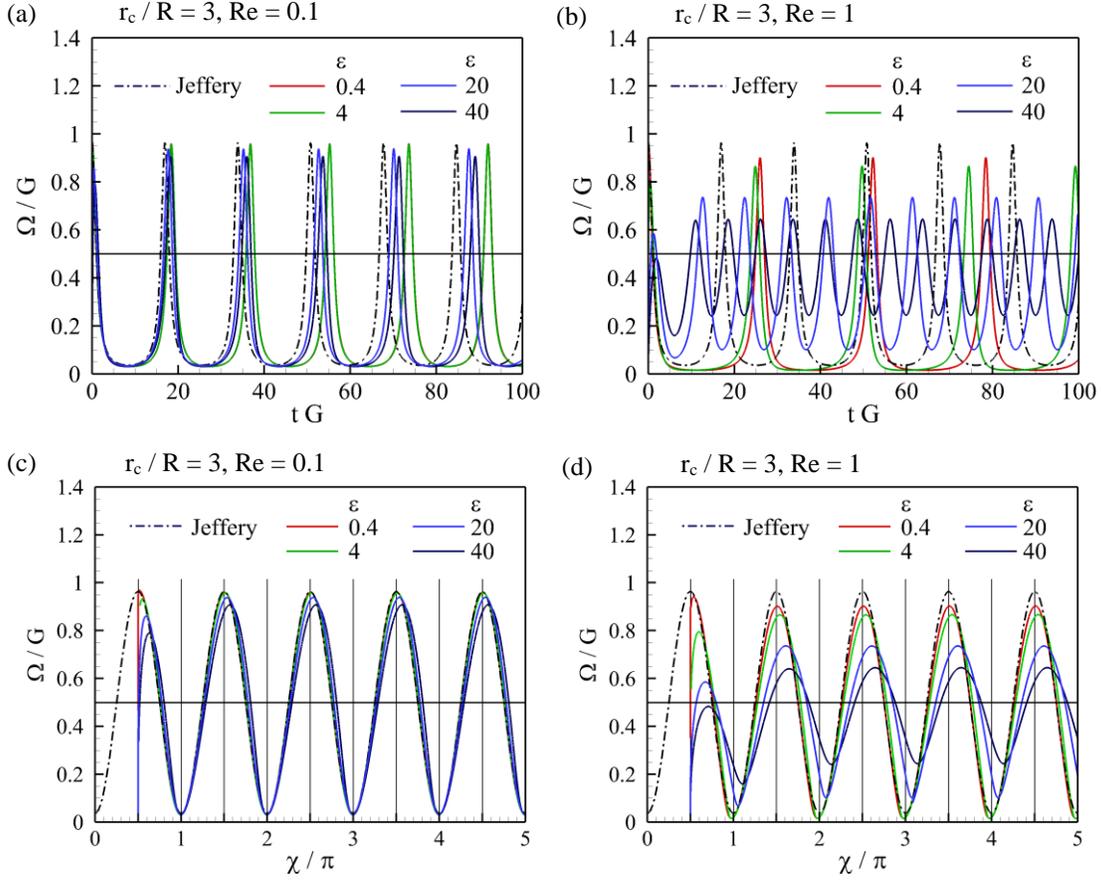

FIGURE 5. Variation of normalized angular velocity ($\Omega/G$) with normalized time ($tG$) (a and b) and rotation angle ($\chi/\pi$) (c and d) of a prolate spheroid with $c/R = 3$ in a simple shear for various density ratios ($\varepsilon$). (a) $\Omega/G$ vs. $tG$ for $Re = 0.1$, (b) $\Omega/G$ vs. $tG$ for $Re = 1$, (c) $\Omega/G$ vs. $\chi/\pi$ for $Re = 0.1$, and (d) $\Omega/G$ vs. $\chi/\pi$ for $Re = 10$. The solid lines are the results of LBM simulation for $Re > 0$ and the dash-dot lines are the prediction of Jeffery's theory for $Re = 0$ (Jeffery 1922).

Figure 6 presents the variations of $\Omega_{min}/G$, $\Omega_{max}/G$, and $TG$ with respect to $\varepsilon$ for a prolate spheroid with $r_c/R = 3$. At a Reynolds number of $Re = 0.1$, the relatively weak fluid and particle inertia only slightly influence $\Omega_{min}/G$, $\Omega_{max}/G$, and $TG$. The rotation period $TG$ is nearly equivalent to that in a Stokes fluid, denoted as $T_S G$. According to Jeffery (1922), this is given by:

$$T_S G = 2\pi(r_a^2 + r_c^2)/r_a r_c \tag{4.1}$$

As the Reynolds number increases to $Re = 1$, the enhanced fluid inertia significantly amplifies the effect of particle inertia, which tends to stabilize the spheroid's angular velocity. As $\varepsilon$ increases from 0.4 to 40, the heightened particle inertia leads to a noticeable decrease in $\Omega_{max}/G$ and a marked increase in $\Omega_{min}/G$, trending towards $\Omega_{min}/G = \Omega_{max}/G = 0.5$ as $\varepsilon$ approaches infinity, as illustrated in figure 6(a) (Lundell & Carlsson 2010; Nilsen & Anderson 2013; Rosén et al. 2015b). The corresponding rotation periods $TG$ as a function of $\varepsilon$ are depicted in figure 6(b). The variation in $\varepsilon$ results in a more pronounced change in $TG$ within the range of $4 \leq \varepsilon \leq 20$. For smaller density ratios ($\varepsilon \leq 4$), particle inertia is weak, and fluid inertia predominantly governs the tumbling, keeping $TG$ close to its value at $\varepsilon = 0$. In contrast, for larger $\varepsilon$ values ($\geq 20$), the sensitivity of $TG$ to $\varepsilon$ diminishes as $\varepsilon$ increases, with the period approaching $4\pi$ as $\varepsilon$ tends towards infinity (Lundell & Carlsson 2010; Nilsen & Anderson 2013; Rosén et al. 2015). The analysis of figures 5 and 6 corroborates the conclusion drawn from figures 2 to 5 that increased particle inertia renders the tumbling of a prolate spheroid more uniform.

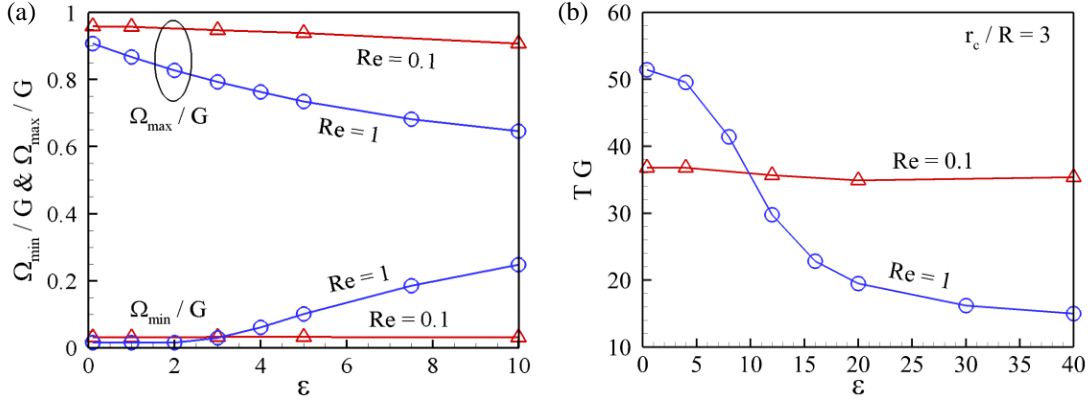

FIGURE 6. Variation of minimum and maximum angular velocities ($\Omega_{min}/G$ and $\Omega_{max}/G$) and rotation period ($TG$) with density ratio ($\varepsilon$). The length of major semi-axis is $r_c/R = 3$. (a) $\Omega_{min}/G$ and $\Omega_{max}/G$ vs. $\varepsilon$, and (b) $TG$ vs. $\varepsilon$.

The tumbling behavior of a prolate spheroid is influenced by its shape, which in this study is represented by the ratio of the major semi-axis length, $r_c/R$. For particles that are nearly spherical ($r_c/R \approx 1$), their behavior and advective transport characteristics closely resemble those of spherical particles. As $r_c/R$ increases, the impact of the non-spherical shape becomes more pronounced. Figure 7 illustrates the dimensionless angular velocity, $\Omega/G$, plotted against dimensionless time, $tG$, and rotation angle, $\chi/\pi$, for $r_c/R$ values of 1.5, 2, 3, and 4. The particle-to-fluid density ratios considered are $\varepsilon = 0.4$ and 40, with Reynolds numbers of $Re = 0.1$ and 1. Beyond the previously discussed effects of fluid and particle inertia, this figure also reveals that a decrease in $r_c/R$ results in a more uniform $\Omega/G$. A detailed analysis is provided in the discussion of figure 8 below.

In figure 8, we present the plots of $\Omega_{min}/G$, $\Omega_{max}/G$, and $TG$ against the ratio $r_c/R$ for particle-to-fluid density ratios $\varepsilon = 0.4$ and 40, and Reynolds numbers $Re = 0.1$ and 1. For spherical particles ($r_c/R = 1$), $\Omega/G$ remains constant over time in the equilibrium state, with values approximately 0.5 for $\Omega/G$ and $4\pi$ for $TG$, showing little dependence on both $Re$ and $\varepsilon$ (see figure 8(a)). As $r_c/R$ increases, the effects of fluid and particle inertia vary under different combinations of $Re$ and $\varepsilon$, leading to distinct trends in $\Omega/G$ variation. A common observation is that with increasing $r_c/R$, $\Omega_{max}/G$ tends to increase while $\Omega_{min}/G$ decreases for each combination of $Re$ and $\varepsilon$. The theoretical limit values are $\Omega_{max}/G = 1$ and $\Omega_{min}/G = 0$ as $r_c/R$ approaches infinity. Among the four combinations depicted, the scenario with $Re = 1$ and $\varepsilon = 40$ exhibits the highest fluid and particle inertia. In this case, $\Omega_{min}/G$ and $\Omega_{max}/G$ differ significantly from the other three scenarios, which are relatively similar to each other. The highest particle inertia results in the smallest difference between $\Omega_{max}/G$ and $\Omega_{min}/G$, thereby achieving the most uniform rotation.

Figure 8(b) provides insights into how the rotation period $TG$ is influenced by the ratio $r_c/R$. Across all combinations of Reynolds number ($Re$) and particle-to-fluid density ratio ($\varepsilon$), an increase in $r_c/R$ results in a longer $TG$. At $Re = 0.1$, the effects of both fluid and particle inertia are weak, even at $\varepsilon = 40$. Consequently, the values of $TG$ for $\varepsilon = 0.4$ and 40 closely align with Jeffery's theoretical predictions for $Re = 0$ for each $r_c/R$ (Jeffery, 1922). When $\varepsilon = 0.4$ and $Re = 1$, the increased fluid inertia significantly elevates $TG$ compared to $Re = 0$, as discussed in figures 2-4. In contrast, For $\varepsilon = 40$ and $Re = 1$, the dominant particle inertia surpasses fluid inertia, primarily influencing the particle's tumbling behavior. This dominance notably raises the minimum angular velocity $\Omega_{min}/G$, thereby substantially reducing the rotation period $TG$. Specifically, as $r_c/R$ increases from 1 to 4, $TG$ rises from 12.57 (approximately $4\pi$) to 16.05, which is considerably lower than in other scenarios.

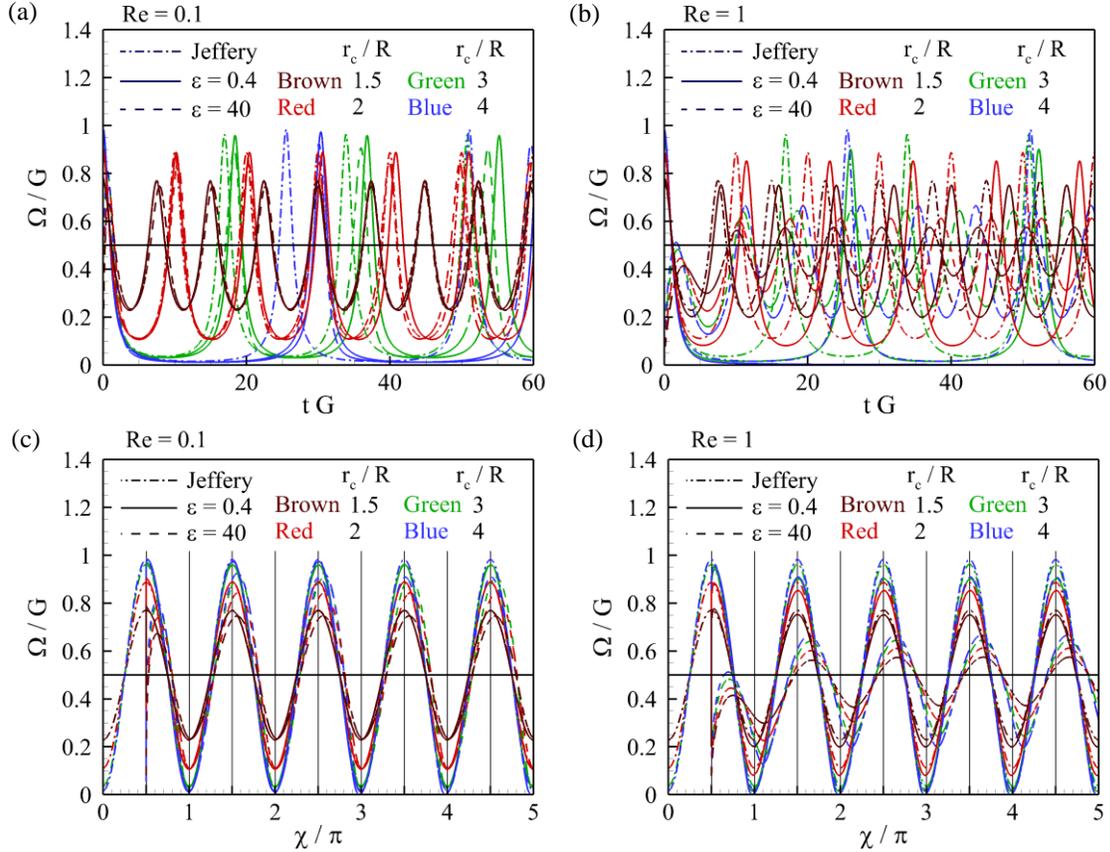

FIGURE 7. Variation of normalized angular velocity ($\Omega/G$) with normalized time ($tG$) and rotation angle ($\chi/\pi$) for various length of major semi-radii ($r_c/R$) and density ratios ($\varepsilon$). (a,b) $\Omega/G$ vs. $tG$, and (c,d) $\Omega/G$ vs. $\chi/\pi$. (a,c) $Re = 0.1$, and (b,d) $Re = 1$. The horizontal line at $\Omega/G = 0.5$ indicates the angular velocity of a sphere at $Re = 0$. Results are compared with the prediction of Jeffery's theory for $Re = 0$ (Jeffery 1922).

Figures 7 and 8 illustrate that, in addition to increasing particle inertia, reducing the ratio $r_c/R$ towards 1 can also lead to a more uniform angular velocity of spheroid tumbling, bringing it closer to the constant angular velocity observed in spherical particles. However, the increased uniformity resulting from a decreased $r_c/R$ differs from that achieved through enhanced particle inertia, particularly in terms of the fluid flow dynamics and advective transport around the particle.

The analysis above concludes that the uniformity of prolate spheroid tumbling is primarily influenced by the interplay between fluid inertia ($Re$), particle inertia ($St$), and the length of the major semi-axis ($r_c/R$) of the spheroid. Fluid inertia predominantly affects uniformity only when particle inertia is weak. In this study, tumbling uniformity encompasses both the amplitude of changes in angular velocity and the disparity in time spans for higher and lower angular velocities. Non-uniform tumbling is characterized by a smaller minimum angular velocity, resulting in a longer period. The uniformity of spheroid tumbling can largely be assessed by the minimum angular velocity ($\Omega_{min}/G$) and the tumbling period ($TG$). Figure 9 presents a regime diagram illustrating the tumbling uniformity of prolate spheroids, depicted by $TG$ and $\Omega_{min}/G$ across a range of $St$ for different $r_c/R$ and $Re$ values. Based on $TG$ and $\Omega_{min}/G$, the regime can be divided into three sub-regimes: I. relatively non-uniform tumbling of slender spheroids with weak particle inertia, II. relatively uniform tumbling of nearly spherical spheroids with weak inertia, and III. relatively uniform tumbling of slender spheroids with stronger inertia. In contrast, the effect of fluid inertia on particle tumbling is less pronounced within the range of Reynolds numbers considered in this paper. The varying degrees of spheroid tumbling uniformity may lead to different flow characteristics around the particle, resulting in distinct advective transport characteristics.

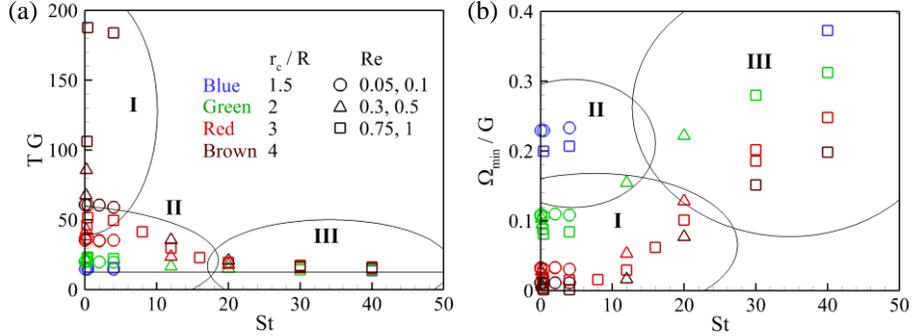

FIGURE 9. Regime diagram divided according to (a) rotation period ($TG$) and (b) minimum angular velocity ($\Omega_{min}/G$). Regime I. relatively non-uniform tumbling of slender prolate spheroid with weak particle inertia, regime II. relatively uniform tumbling of nearly spherical prolate spheroids with weak inertia, regime III. Relatively uniform tumbling of slender prolate spheroids with strong inertia.

## *4.2 Characterization of Advective Scalar Transport*

To explore the potential impact of the uniformity of prolate spheroid tumbling on flow and transport characteristics around the particles, we categorize the parameter space ($1 \leq r_c/R \leq 4$, $Re \leq 1$, and $0.4 \leq \varepsilon \leq 40$) into three distinct regimes, as illustrated in figure 9. In Regime II, the flow patterns generated by the tumbling of nearly spherical particles resemble the uniform rolling of a sphere, a phenomenon that has been extensively studied in recent years (Wang 2019). This section focuses on characterizing the fluid flow and advective transport induced by both non-uniform and relatively uniform tumbling of slender prolate spheroids. For this analysis, we select a prolate spheroid with $r_c/R = 3$ as a representative example. The Reynolds number is set at $Re = 1$, and we consider two particle-to-fluid density ratios, $\varepsilon = 0.4$ and 40, which correspond to weak and strong particle inertia, respectively.

The advective transport of a passive scalar is illustrated through the trajectories of fluid particles that carry the scalar. In the context of unsteady flows induced by particle tumbling, neither instantaneous nor time-averaged flow fields adequately capture the scalar transport process. Inspired by the use of streak lines in flow visualization (Lane 1993), we employ continuously released virtual micro tracers around the spheroid to discern flow and transport characteristics. These tracers are released at predetermined time intervals from specific locations around the spheroid. In the simulation of fluid flow, a time interval of twenty steps is used for lower Reynolds numbers, while ten steps are applied for higher Reynolds numbers. These intervals are sufficiently small to accurately capture the flow patterns. Importantly, the presence of micro tracers does not affect the flow evolution, as their velocities match the fluid velocities at each position. The new positions of the micro tracers at each time interval are computed using the fourth-order Runge-Kutta method.

We first consider the case where $\varepsilon = 0.4$, indicating weak particle inertia and more non-uniform spheroid tumbling. Figure 10 illustrates the instantaneous streaklines of micro tracers around a prolate spheroid with $r_c/R = 3$ and $\varepsilon = 0.4$ at $Re = 1$. The rotation angle is $\chi = 5\pi/8$. Micro tracers are released from various locations: along a vertical line upstream of the spheroid on the central x-z plane (figure 10(b)), along two horizontal lines perpendicular to the $x$-$z$ plane upstream of the spheroid (figure 10(c)), and at two points on the lateral side of the spheroid (figures 10(d) and 10(e)). The specific release locations are detailed in the caption of figure 10. The streaklines from these different release points combine to form a complex pattern, as depicted in figure 10(a).

On the central $x$-$z$ plane, as shown in figure 10(b), the motion of the tracers remains confined to this plane. Micro tracers released near the flow axis at coordinates $x = -6R$, $y = 0$, and $z = r_c/8$ initially move towards the spheroid. At a certain point, they reverse direction and move away, forming a recirculating wake near the flow axis, a phenomenon commonly observed with spherical particles. For tracers released slightly higher at $x = -6R$, $y = 0$, and $z = r_c/4$, their streaklines break into several segments based on the release time. Some segments pass over the spheroid and continue downstream, either to the right with the upper flow or to the left with the lower flow. Other segments become trapped by the spheroid, forming an open fluid layer on its surface. This fluid layer separates the passing fluid from the spheroid surface and rotates with it. Tracers released at even higher positions, where $z > r_c/4$, move towards the spheroid, converge near the upper pole, and then disperse as they travel downstream. The behavior of micro tracers on the central x-z plane illustrates the fluid flow within a specific lateral range near this plane, which is crucial for the transport of passive scalars.

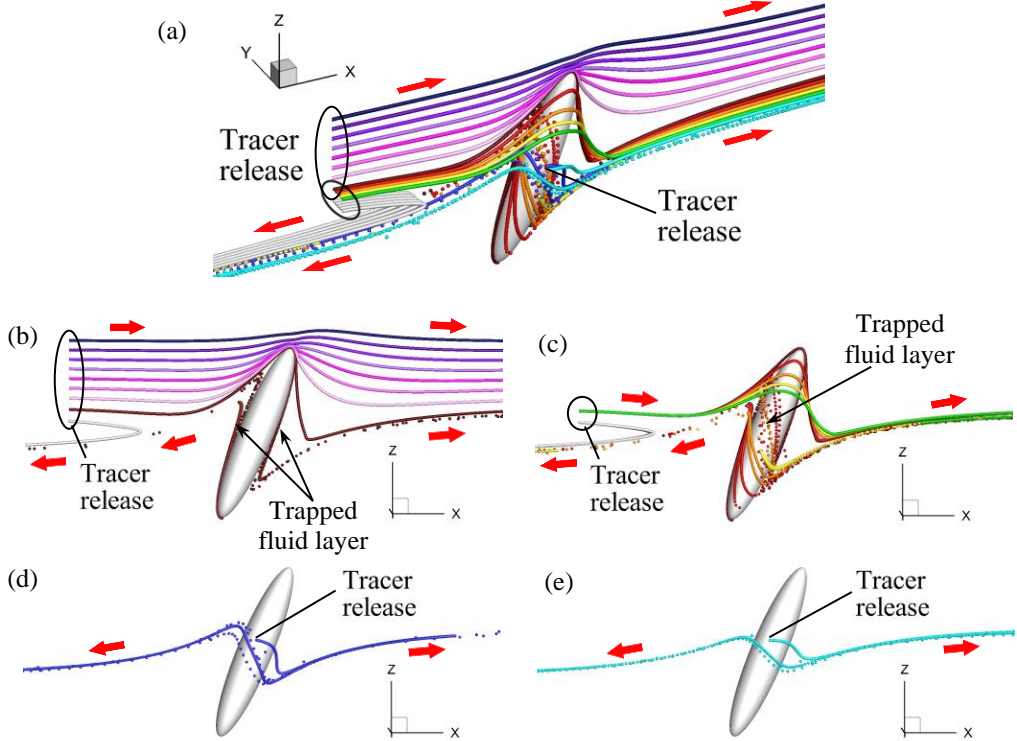

FIGURE 10. Instantaneous patterns of streaklines of micro tracers released at different locations near a prolate spheroid with $r_c/R = 3$ and $\varepsilon = 0.4$ for $Re = 1$. The rotation angle is $\chi = 5\pi/8$. The continuous lines are composed of discrete tracers. (a) Overall pattern of micro tracers released from a vertical line on the central $x$-$z$ plane, $x = -6R$, $y = 0$, $z = mr_c/10$, where $m = 4, \cdots, 10$, from two horizontal lines perpendicular to the $x$-$z$ plane, $x = -6R$, $y = nr_a/4$, $z = r_c/8$ and $r_c/4$, where $n = 0, \cdots, 4$, and from two points on lateral side, $x = 0$, $y = 6r_a/5$ and $8r_a/5$, $z = r_c/8$, (b) tracers released on the central $x$-$z$ plane, including those from the vertical line and the intersection points of two perpendicular lines on the central $x$-$z$ plane, (c) tracers released from two lines perpendicular to the central $x$-$z$ plane, (d) tracers released from a point closer to the lateral side of the spheroid ($y = 6r_a/5$), and (e) from a point slightly farther from the lateral side ($y = 8r_a/5$).

Figure 10(c) illustrates the streaklines of micro tracers released from two horizontal lines that are perpendicular to the central x-z plane. The streaklines originating from the lower line at coordinates $x = -6R$ and $z = r_c/8$ initially move towards the spheroid before reversing direction, creating a recirculating wake. In contrast, the streaklines from the upper line at $x = -6R$ and $z = r_c/4$ wrap around the lateral side of the spheroid. Near the central x-z plane, these streaklines fragment into multiple segments. Some segments proceed directly downstream to the right, while others become trapped along the spheroid's side, forming an open fluid layer. Meanwhile, streaklines positioned farther from the x-z plane continue directly downstream to the right.

Figures 10(d) and 10(e) display the streaklines of micro tracers released from two positions on the lateral side of the spheroid. One position is closer to the spheroid's side at coordinates $x = 0$, $y = 6r_a/5$, $z = r_c/8$, while the other is slightly farther away at $x = 0$, $y = 8r_a/5$, $z = r_c/8$. Both streaklines exhibit similar patterns, characterized by their division into several segments that move in different directions. The movement of these micro tracers and the resulting streaklines offer valuable insights into the mechanisms underlying advective scalar transport.

Figure 11 presents the instantaneous patterns of scalar concentration $\phi$ and representative streaklines of micro tracers surrounding the spheroid for a Schmidt number of $Sc = 100$. In these patterns, regions of high concentration extend downstream, either to the upper right or the lower left, following the streaklines. This observation indicates that the advective transport of the scalar is driven by the movement of fluid particles traversing the spheroid's surface. The subsequent analysis will delve into the fluid flow dynamics and the consequent transport process of the passive scalar.

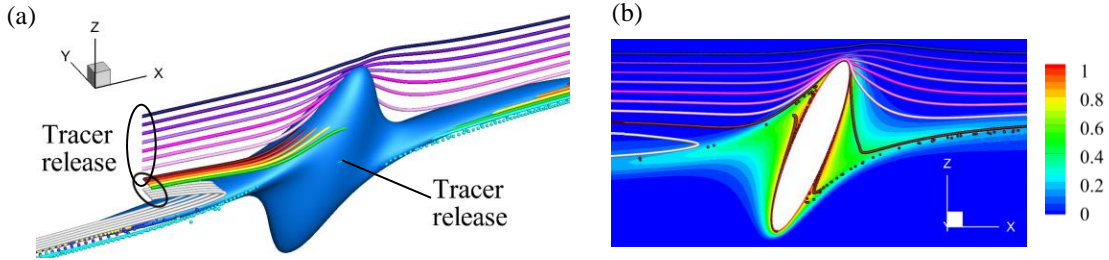

FIGURE 11. Instantaneous patterns of the concentration of passive scalar around a prolate spheroid of $r_c/R = 3$ and $\varepsilon = 0.4$, with typical streaklines of micro tracers released at different positions. Reynolds number $Re = 1$, and Schmidt number $Sc = 100$. The rotation angle $\chi = 5\pi/8$. The positions where micro tracers are released are the same as those shown in FIG. 10. (a) 3D iso-surfaces with $\phi = 0.2$, and (b) 2D iso-contours on the central x-z plane.

Figure 12 depicts the evolution of streaklines of micro tracers released upstream of the spheroid on the central x-z plane, alongside the 3D and 2D patterns of scalar concentration $\phi$, during half a cycle of tumbling. Initially, micro tracers with zero scalar concentration are released upstream and follow the streaklines toward the spheroid. As the rotation angle $\chi$ increases from 0 to $4\pi/8$, the streaklines converge near the spheroid's upper pole, where the micro tracers acquire passive scalar from the spheroid surface through diffusion. As $\chi$ progresses from $4\pi/8$ to $\pi$, the micro tracers on various streaklines that converge near the upper pole move downstream, forming a material line with elevated scalar concentration, originating from the upper pole. This material line is referred to as a scalar line (SL) due to its increased scalar concentration. The downward motion of the spheroid's upper pole drags the scalar line downward, sweeping through the right wake area. Concurrently, fluid particles on the scalar line move downstream to the right, transporting passive scalar through advection. When the spheroid's orientation becomes horizontal, with $\chi$ reaching $\pi$, the newly formed scalar line merges with scalar line clusters generated in previous cycles. Throughout the cycle, some fluid from the flow layer on the spheroid surface detaches and transports passive scalar downstream via the clustered scalar lines. In the lower half area around the spheroid, a newly generated scalar line sweeps upward through the left downstream area, merging into the scalar line clusters on the left. The generation and movement of these scalar lines constitute the advective scalar transport mechanism near the central x-z plane. This generation of scalar lines is primarily due to the anisotropy of the particle shape. For isotropic particle shapes concerning the axis of rotation, such as spherical particles and rotating oblate spheroids, scalar lines are not generated (Wang et al. 2019; Wang et al. 2023).

We are interested in understanding how fluid particles enter and exit the fluid layer during the tumbling of the spheroid. Figure 13 illustrates the movement of characteristic fluid particles, represented by micro tracers, on two streaklines: one close to the spheroid surface (released at $x = -20R, y = 0, z = 5r_c/20$) and the other farther from the spheroid (released at $x = -20R, y = 0, z = 15r_c/20$). These characteristic micro tracers are labeled from 1 to 7 and identified by different symbols. Due to the periodic nature of the flow field, the movements of micro tracers released at the same rotation angle of the spheroid in different cycles are identical, and thus they are assigned the same numbers and symbols. Consequently, the segment of streaklines between the two points marked as '1' at $\chi = \pi / 8$ encompasses all tracers released within a half cycle of spheroid tumbling. These characteristic tracers divide the streamlines into several segments with distinct behaviors.

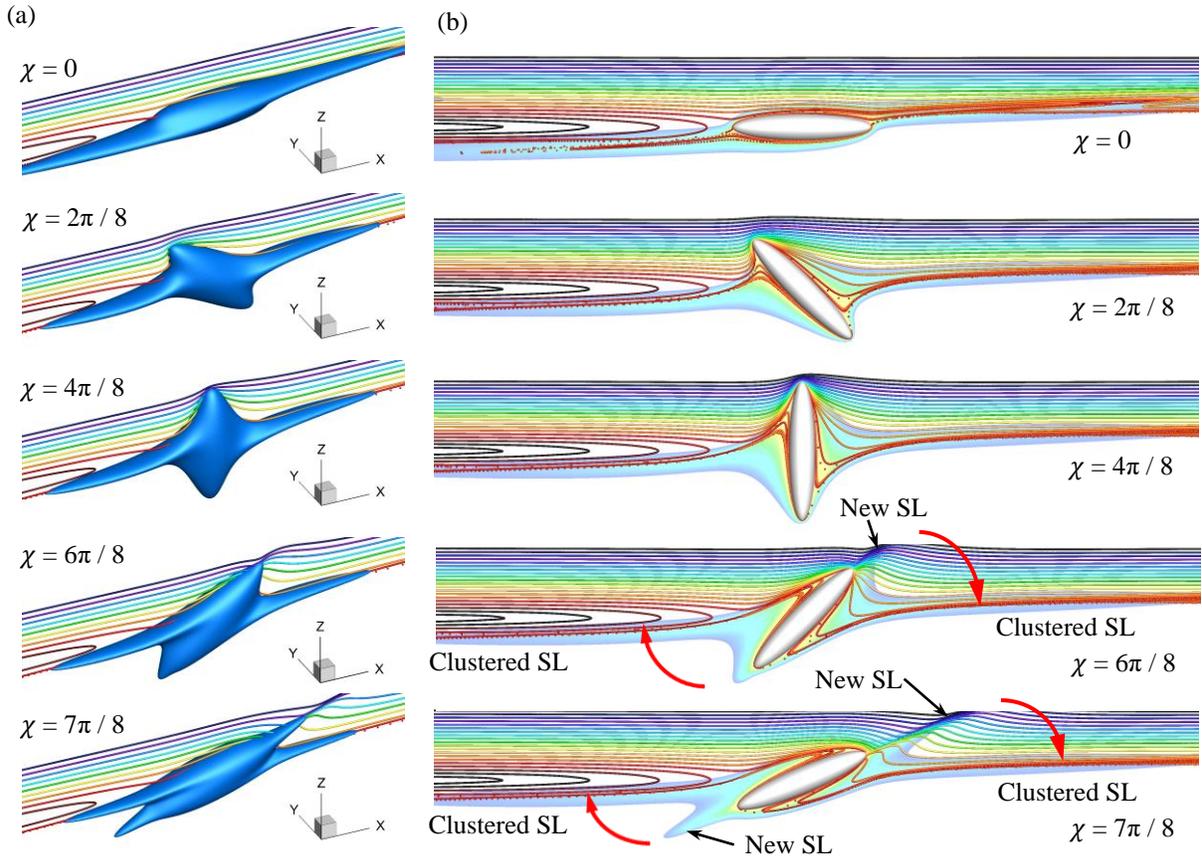

FIGURE 12. Evolution of the streaklines of micro tracers released upstream of the spheroid on the central x-z plane, at $x = -20R, y = 0, z = mr_c/20$, where $m = 1, \cdots, 20$, together with the 3D iso-surfaces and 2D iso-contours of concentration of passive scalar. $r_c/R = 3$, $\varepsilon = 0.4$, $Re = 1$, and $Sc = 100$. (a) 3D representation with iso-surfaces at $\phi = 0.25$, and (b) 2D representation on the central x-z plane. Iso-contours are cutoff at $\phi = 0.15$. SL: scalar lines defined as material lines with concentrated passive scalar.

On the streakline close to the spheroid, tracer '1' (Δ) serves as the leading point of the streakline segments considered in this analysis. During the tumbling process, the upper pole of the spheroid severs the streakline at tracer '2' ( | ), allowing the segment '1-2' to travel directly downstream to the right, while the segments following tracer '2' ( | ) become trapped by the tumbling spheroid within the fluid layer on its surface. Tracer '2' ( | ) marks the transition point between segments moving in opposite directions relative to the spheroid's pole, thus remaining stationary with respect to the spheroid. Tracer '3' (○) leads the trapped streakline segments, guiding segments '2-3' and '3-4' along the spheroid's surface in tandem with its tumbling motion. As tracer '4' (●) nears the spheroid, it attaches to the polar area and rotates with the spheroid, due to the opposing directions of the streakline segments before and after it. Led by tracer '5' (♦), the streakline segments following tracer '4' (●), including '4-5', '5-1', and '1-2', proceed downstream to the right until the streakline is severed by the spheroid's pole at tracer '2' in the subsequent cycle. Essentially, tracer '1' (Δ) is not as unique as the other tracers. Observing the streakline patterns from $\chi = 15\pi/16$ to $9\pi/8$, it is evident that the evolution of segments trapped in the fluid layer (below the spheroid) mirrors that of segments approaching the spheroid (above the spheroid). This indicates that the trapped streakline segments will undergo the same process as those approaching the spheroid, with one part moving directly downstream and the other remaining trapped in the fluid layer. This cycle repeats indefinitely, meaning some tracers will never leave the spheroid. The fluid particles represented by the streakline segments trapped in the fluid layer maintain close contact with the spheroid, allowing them to acquire passive scalar from its surface. Upon leaving the spheroid, they follow the clustered scalar lines depicted in figure 12, with an elevated scalar concentration, thereby transporting passive scalar downstream.

On the streakline slightly farther from the spheroid (released at $x = -20R$, $y = 0$, $z = 15r_c/20$), tracers '6' and '7' initially converge at the spheroid's upper pole, gaining passive scalar from its surface. They then depart from the spheroid along a newly generated scalar line with a higher scalar concentration, transporting the passive scalar downstream. Ultimately, the newly generated scalar line descends and merges into the scalar line cluster.

Figure 13(b) illustrates the trajectories of the characteristic micro tracers. Notably, tracers '2' ( | ) and '4' (●) are the only ones attached to the poles of the spheroid, rotating along with it. In contrast, the other characteristic tracers, do not rotate with the spheroid. The same is true for the non-characteristic tracers. Only those micro tracers in close proximity to tracers '2' ( | ) and '4' (●) become trapped in the fluid layer and follow the spheroid's rotation. This behavior is primarily due to the non-uniform angular velocity associated with the spheroid's tumbling motion.

On the lateral sides of the spheroid, the behavior of micro tracers varies distinctly. Figure 14 illustrates the movement of micro tracers released at the position $x = 0$, $y = 6r_a/5$, $z = r_c/8$. Several characteristic tracers, labeled from 1 to 4, divide the streakline into multiple segments. Initially, segment '1-2' turns left with the spheroid, then detaches and moves downstream to the left. Tracer '2' temporarily remains near the spheroid, moving in tandem with it, while tracer '3' behaves similarly on the opposite side. Tracer '4' leads segment '4-3' and the segment between itself and the newly released tracer '1' downstream to the right. When $\chi = 20\pi / 16$ and $\chi = 24\pi / 16$, tracers '2' and '3' are positioned symmetrically relative to the spheroid's center, indicating that the segment between them traverses the spheroid's center. In subsequent cycles, tracers '2' and '3' move away from the spheroid, stretching the segment while maintaining a portion that passes over the spheroid, causing some tracers to remain with the spheroid for multiple cycles. The iso-contours of scalar concentration on the central *x-z* plane suggest that micro tracers released on the lateral sides also transport passive scalar downstream along clustered scalar lines. Figure 14(b) shows the trajectories of the characteristic tracers, indicating that none of them rotate with the spheroid due to its non-uniform tumbling. As demonstrated in figure 10, streaklines released on the lateral side exhibit similar patterns to those released upstream of the spheroid but off the central $x$-$z$ plane. Consequently, the tracer movement and scalar transport described in figure 14 apply to all areas on the sides of the spheroid.

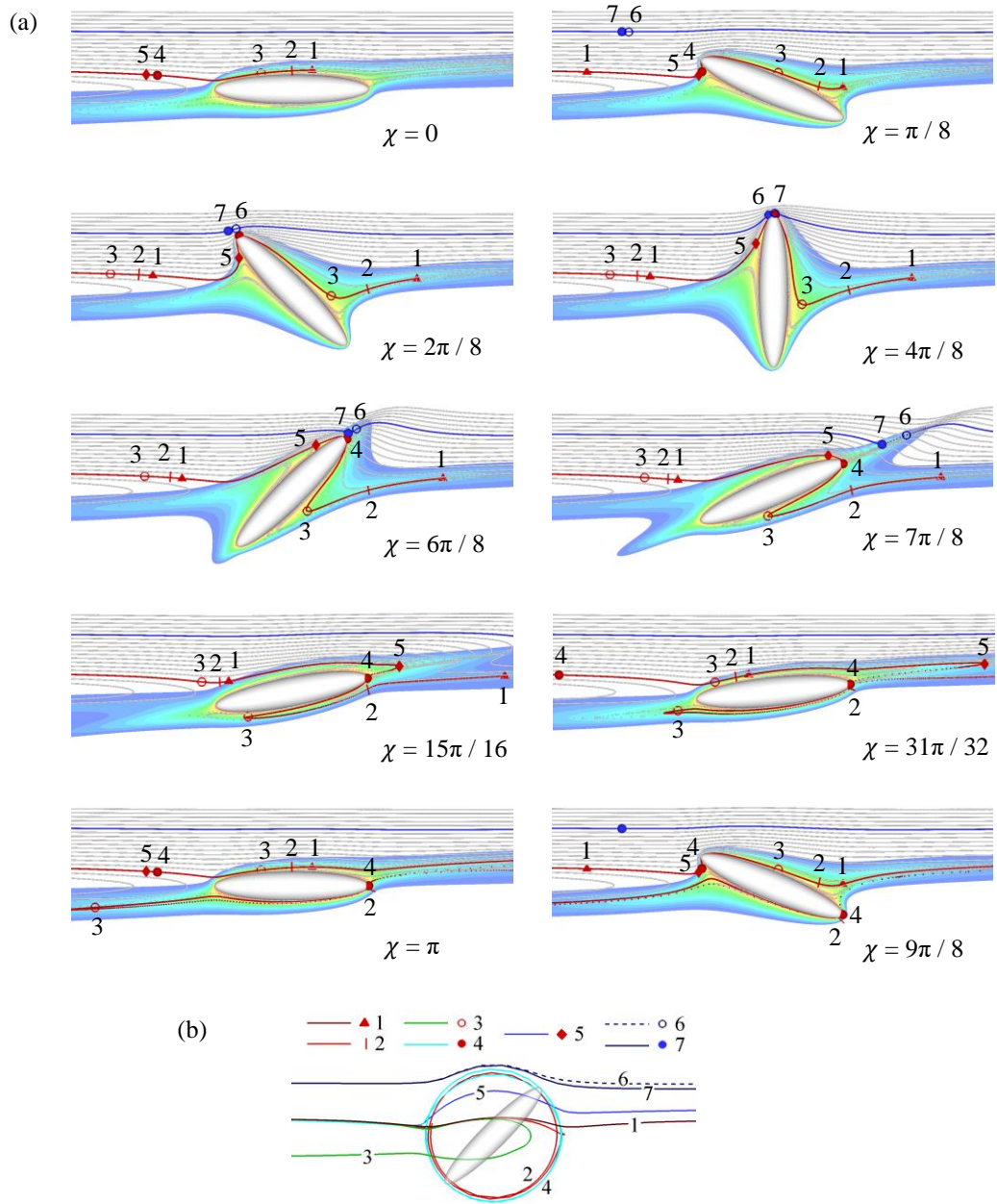

FIGURE 13. Movement of fluid particles represented by micro tracers released at $x = -20R$, $y = 0$, $z = 5r_c/20$ and $15r_c/20$ on the central $x$-$z$ plane. $r_c/R = 3$, $\varepsilon = 0.4$, $Re = 1$, and $Sc = 100$. (a) Streaklines of micro tracers with several characteristic micro tracers marked by numbers, together with the 2D iso-contours of scalar concentration on the central plane, and (b) trajectories of the characteristic micro tracers. The 2D iso-contours are cutoff at $\phi = 0.15$.

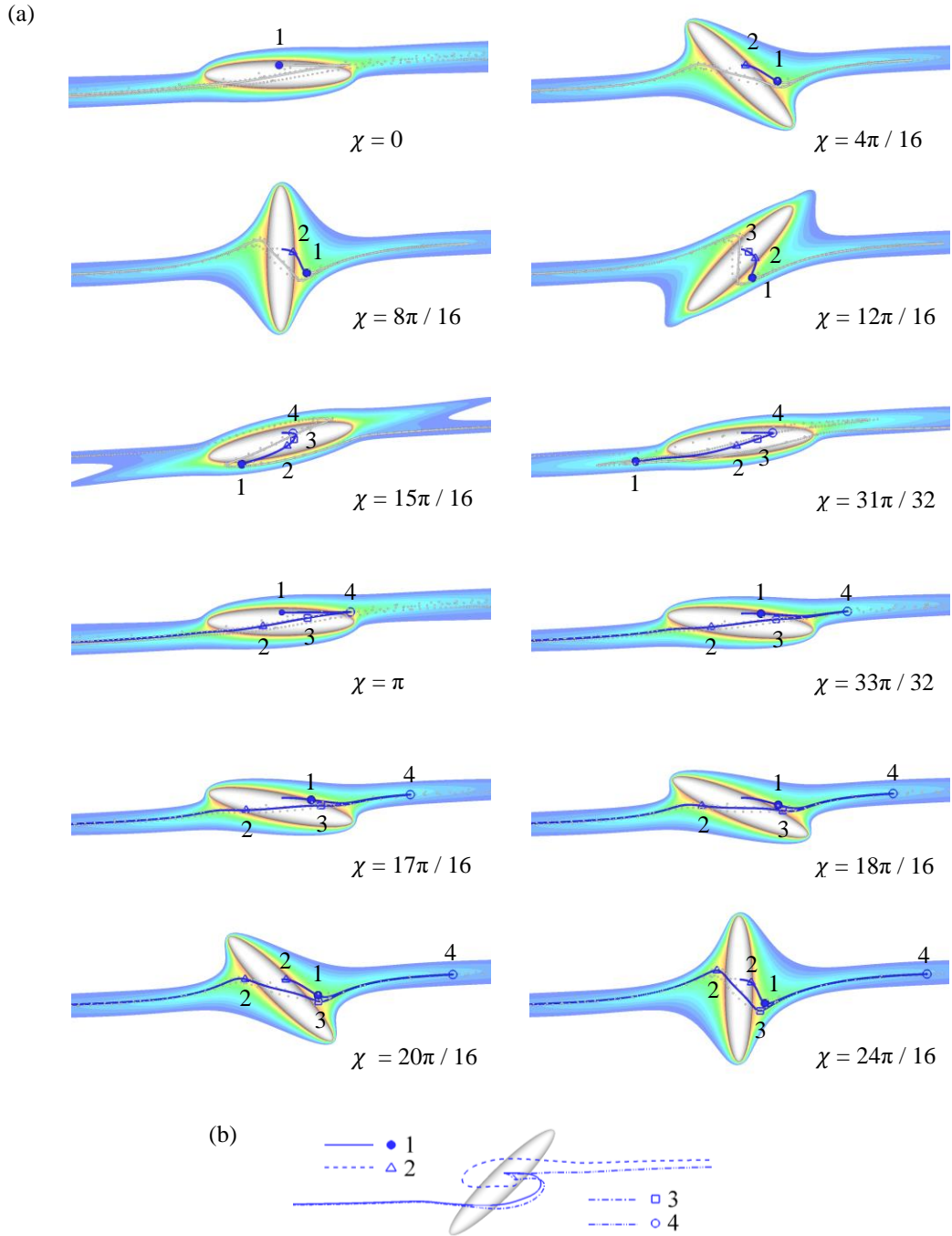

FIGURE 14. Movement of fluid particles represented by micro tracers released on the lateral side at $x = 0$, $y = 6r_a/5$, $z = r_c/8$ of a prolate spheroid with $r_c/R = 3$ and $\varepsilon = 0.4$ for $Re = 1$ and $Sc = 100$. (a) Streaklines of micro tracers with several characteristic micro tracers marked by numbers, together with the evolution of 2D iso-contours of scalar concentration on the central $x$-$z$ plane, and (b) trajectories of the characteristic micro tracers. The 2D iso-contours are cutoff at $\phi = 0.15$.

When the particle-to-fluid density ratio increases to $\varepsilon = 40$, the spheroid's tumbling becomes more uniform due to enhanced particle inertia, resulting in a distinct fluid flow pattern compared to $\varepsilon = 0.4$. Figure 15 illustrates the instantaneous streaklines of micro tracers around a prolate spheroid with a ratio of $r_c/R = 3$ and $\varepsilon = 40$ at $Re = 1$, with a rotation angle of $\chi = 5\pi/8$. The micro tracers are released from the same position as in figure 10. Compared to the patterns observed at $\varepsilon = 0.4$, the increased particle inertia introduces several notable differences. Firstly, the reduction in $(\Omega/G)_{max}$ amplifies the velocity difference between the spheroid surface and the surrounding fluid when the spheroid's major axis is vertical. This causes the streaklines trapped in the fluid layer to detach from the spheroid surface on the leeward side, as depicted in figure 15(b). Secondly, the more uniform tumbling of the spheroid induces the fluid on the lateral sides to rotate with it. Under the influence of centrifugal force, the micro tracers released on the side spiral outward, increasing their radial coordinate. Tracers released closer to the spheroid ($x = 0$, $y = 6r_a/5$, $z = r_c/8$) exhibit an S-shaped twisted spiral pattern, as shown in figure 15(d), while those released slightly further away ($x = 0$, $y = 8r_a/5$, $z = r_c/8$) display a normal outward spiral pattern, as shown in figure 15(e). Flow characteristics around a sphere (Subramanian & Koch 2006a,b; Wang & Brasseur 2019) suggest that these two spiral flows originate from a region around the axis of rotation farther from the spheroid. They recirculate on the spheroid sides and then travel downstream along the clustered scalar lines near the central x-z plane. Thirdly, the micro tracers released upstream of the spheroid ($x = -6R$, $y =$ $nr_a/4, z = r_c/4$, where $n = 1, \cdots, 4$) are separated from the spheroid by the spiral flows on the side and do not reciprocate across the side surface, as shown in figure 15(c). These changes in flow patterns significantly alter the patterns of scalar transport.

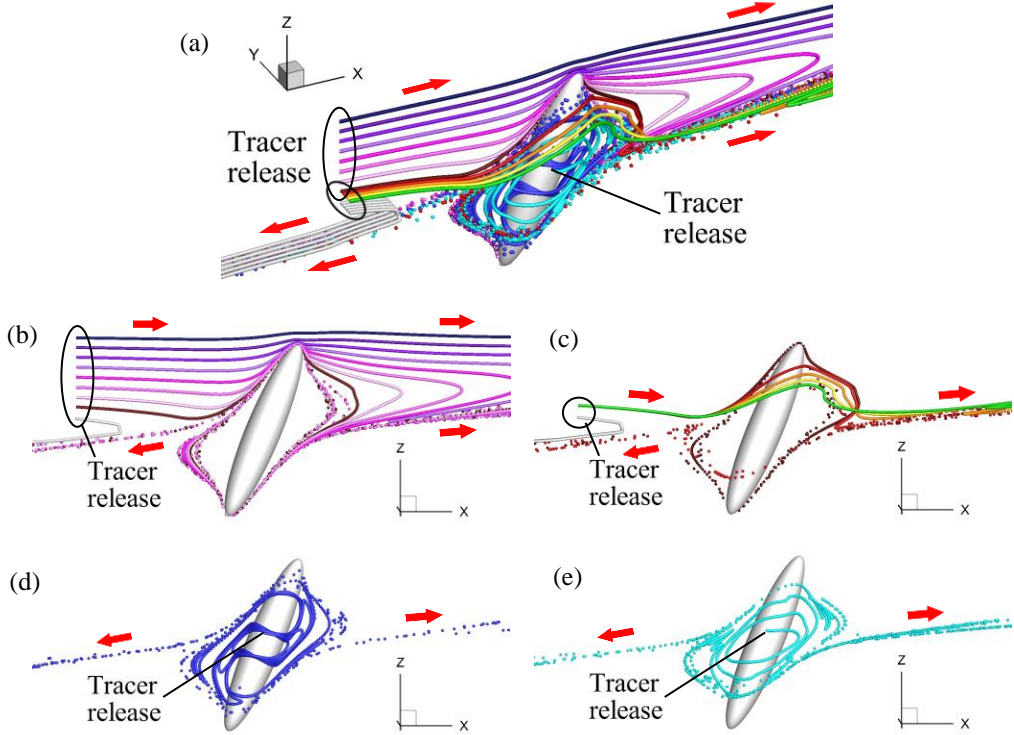

FIGURE 15. Instantaneous patterns of streaklines of micro tracers released at different locations near a prolate spheroid with $r_c/R = 3$ and $\varepsilon = 40$ for $Re = 1$. The rotation angle is $\chi = 5\pi/8$. The continuous lines are composed of discrete tracers. (a) Overall pattern of micro tracers released from a vertical line on the central x-z plane, $x = -6R$, $y = 0$, $z = mr_c/10$, where $m = 4, \cdots, 10$, from two horizontal lines perpendicular to the x-z plane, $x = -6R$, $y = nr_a/4$, $z = r_c/8$ and $r_c/4$, where $n = 0, \cdots, 4$, and from two points on lateral side, $x = 0$, $y = 6r_a/5$ and $8r_a/5$, $z = r_c/8$, (b) tracers released on the central x-z plane, including those from the vertical line and the two intersection points of perpendicular lines on the central x-z plane, (c) tracers released from two lines perpendicular to the central x-z plane, (d) tracers released from a point closer to the lateral side of the spheroid ($y = 6r_a/5$), and (e) tracers relaced from a point slightly farther from the lateral side ($y = 8r_a/5$).

Figure 16 illustrates the instantaneous patterns of scalar concentration $\phi$ around a spheroid with a radius ratio $r_c/R = 3$ and a density ratio $\varepsilon = 40$ at Reynolds number $Re = 1$ and Schmidt number $Sc = 100$. The typical streaklines of micro tracers are released from the same positions as in figure 15, with the rotation angle set at $\chi = 5\pi/8$. In comparison to the patterns observed at $\varepsilon = 0.4$ (as shown in figure 11), a notable difference is the detachment of streaklines within the fluid layer, which causes the region of higher scalar concentration to expand significantly downstream on the leeward side. This alteration could potentially impact the efficiency of scalar transport.

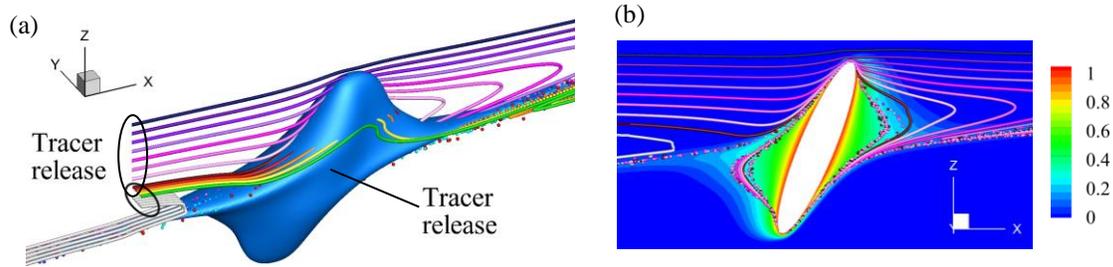

FIGURE 16. Instantaneous patterns of the concentration of passive scalar around a prolate spheroid of $r_c/R = 3$ and $\varepsilon = 40$, with typical streaklines of micro tracers released at different positions. Reynolds number $Re = 1$, and Schmidt number $Sc = 100$. The rotation angle $\chi = 5\pi/8$. The positions where micro tracers are released are the same as those shown in FIG. 15. (a) 3D iso-surfaces with $\phi = 0.2$, and (b) 2D iso-contours on the central $x$-$z$ plane.

The advective transport of passive scalar is demonstrated by the evolution of streaklines of micro tracers. Figure 17 depicts the streaklines of micro tracers released upstream of the spheroid on the central $x$-$z$ plane, alongside the patterns of scalar concentration $\phi$, during half a cycle of spheroid tumbling. Similar to the scenario with $\varepsilon = 0.4$, streaklines originating from upstream converge at the upper pole of the spheroid, subsequently forming a scalar line that extends downstream with an elevated scalar concentration. This newly formed scalar line, drawn by the upper pole of the spheroid, sweeps downward through the wake region and merges with the clustered scalar lines generated in previous cycles. Throughout this process, the micro tracers travel downstream along the scalar line, effectively transporting passive scalar away from the spheroid. Concurrently, some micro tracers trapped in the fluid layer on the spheroid's surface detach and travel downstream along the clustered scalar lines, further facilitating the downstream transport of passive scalar.

Figure 18 illustrates the evolution of two streaklines of micro tracers released on the lateral side at positions $x = 0$, $y = 6r_a/5$ and $8r_a/5$, $z = r_c/8$, along with the patterns of scalar concentration. The streakline closer to the spheroid forms a twisted spiral, while the one farther away takes the shape of a normal outward spiral. These spiral streaklines, along with the fluid detached from the leeward side of the spheroid surface, extend downstream on the leeward side, forming a tail in the wake area. This tail travels downstream in conjunction with the scalar lines on the central $x$-$z$ plane. Simultaneously, some segments of the spiral streaklines break away from the spirals and proceed downstream along the clustered scalar lines. These two modes of downstream movement of micro tracers illustrate the advective scalar transport facilitated by the fluid flow on the lateral sides.

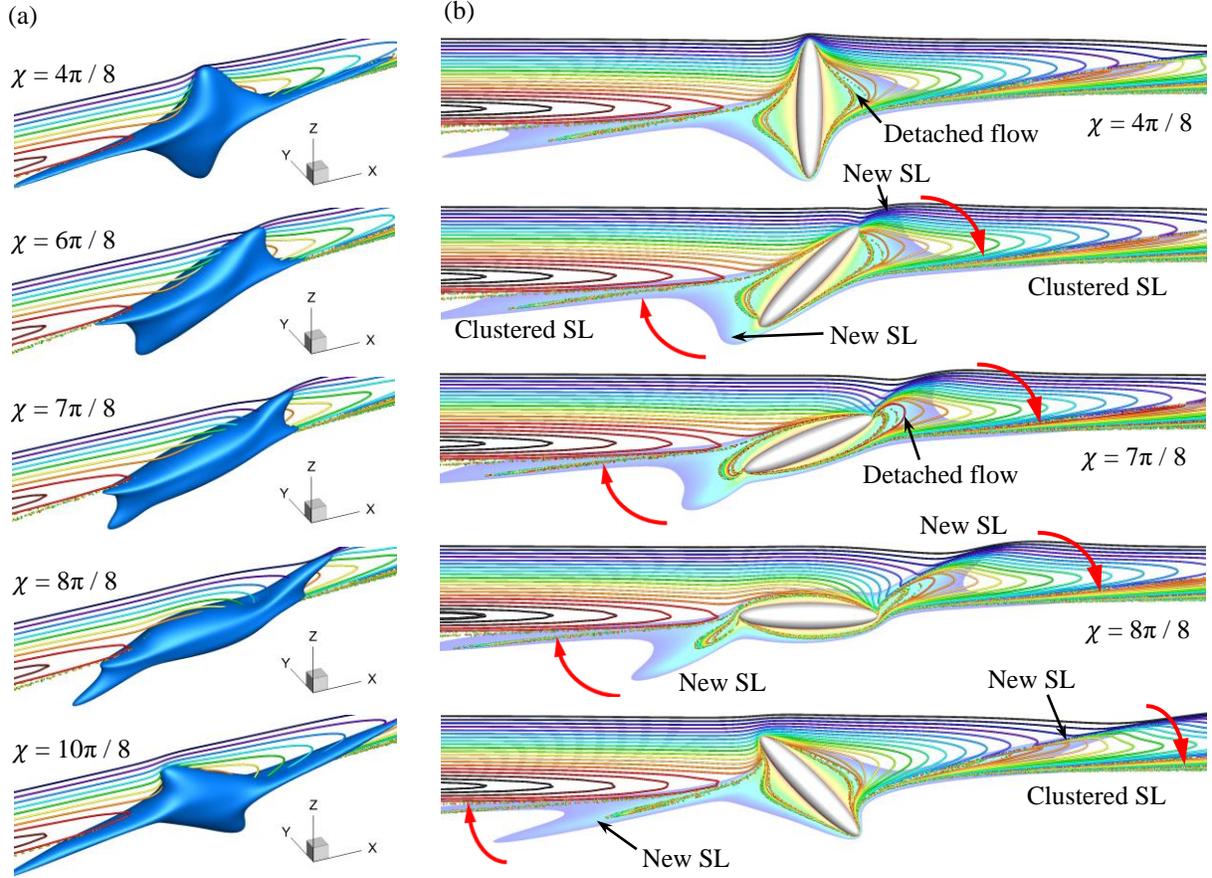

FIGURE 17. Evolution of the streaklines of micro tracers released upstream of the spheroid on the central *x-z* plane, $x = -20R$, $y = 0$, $z = mr_c/20$, where $m = 1, \cdots, 20$, together with the 3D iso-surfaces and 2D iso-contours of the concentration of passive scalar. $r_c/R = 3$, $\varepsilon = 40$, $Re = 1$, and $Sc = 100$. (a) 3D representation with iso-surfaces at $\phi = 0.25$, and (b) 2D representation on the central *x-z* plane. Iso-contours are cutoff at $\phi = 0.15$.

To gain a more detailed understanding of scalar transport, the motion of specific fluid particles is analyzed. Figure 19 illustrates the movement of fluid particles, represented by several characteristic micro tracers, along two streaklines: one close to the spheroid surface (released at $x = -20R$, $y = 0$, $z = 5r_c/20$) and the other farther away (released at $x = -20R$, $y = 0$, $z = 15r_c/20$). The characteristic micro tracers are numbered from 1 to 5 and are distinguished by different symbols. These tracers divide the streamlines into segments, each exhibiting distinct behaviors.

On the streakline near the spheroid, tracer '1' (●) attaches to the spheroid's pole after being released from upstream, as the segments before and after it move in opposite directions. A new tracer '1' (●), released after half a rotation period, attaches to the opposite pole. The streakline segments released during this period ('1-2-3-1') form an arc with both ends anchored to the spheroid poles. This arc moves downstream with the fluid to the right, causing the streaklines to detach from the spheroid surface. When the spheroid is horizontal ($\chi = 8\pi / 8$), the right pole cuts the streaklines at tracer '2' ( | ), causing it to attach to the pole and rotate with the spheroid. The segment before tracer '2' ( | ), '1-2', becomes trapped in the fluid layer on the spheroid surface and rotates to the left. The segments after tracer '2' ( | ), '2-3' and '3-1', travel downstream to the right, guided by tracer '3' (○). On the opposite side, the trapped segment '1-2' forms a similar shape to the segments '1-2-3-1' on the right ($\chi = 10\pi / 8$ and $12\pi / 8$). Consequently, the trapped segment '1-2' undergoes the same cutting process, with one segment moving downstream and the other remaining in the fluid layer, repeating indefinitely.

On the streakline slightly away from the spheroid, tracers '4' and '5' first converge near the upper pole and then travel downstream along the scalar line, as depicted in the figure. The trajectories of these characteristic micro tracers are shown in figure 19(b). Tracers attached to the spheroid ('1' and '2') rotate with it, while the others ('3', '4', and '5') move directly downstream. The motion of these characteristic micro tracers and streakline segments illustrates how passive scalar is transported downstream from the spheroid's surface near the central $x$-$z$ plane.

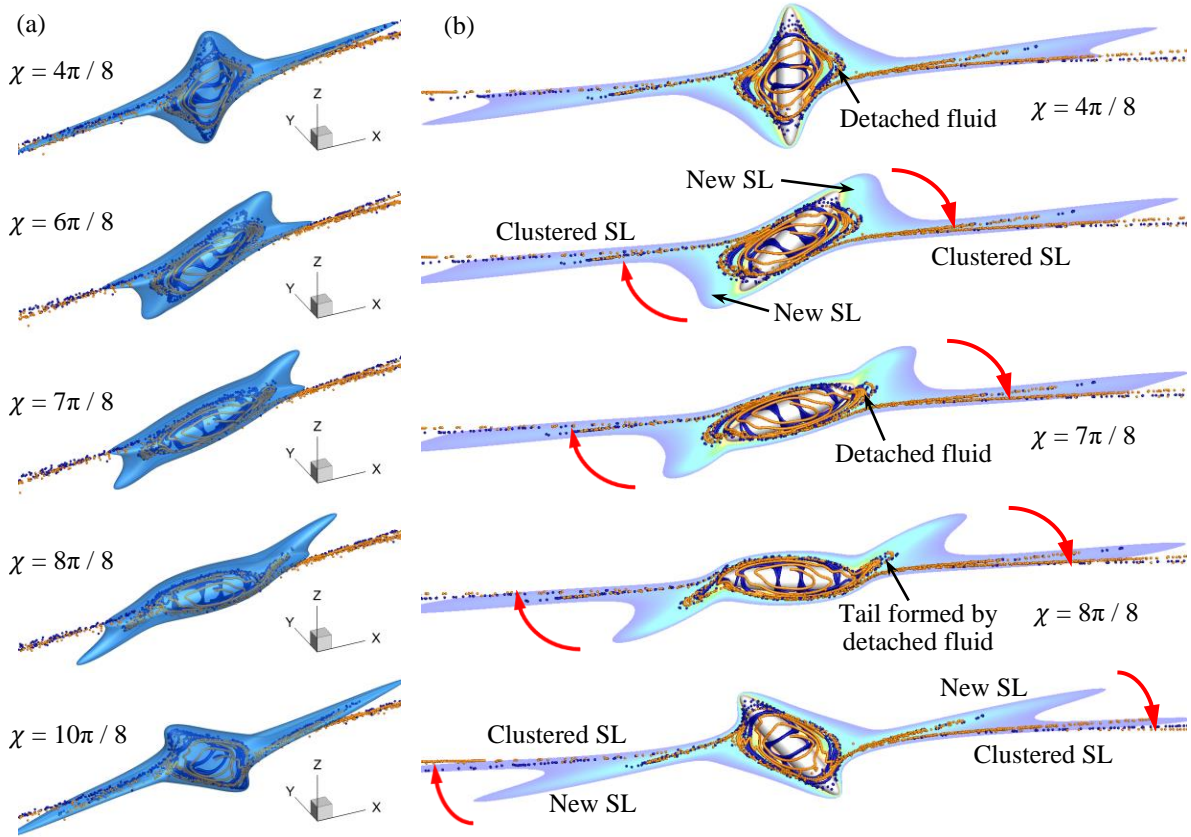

FIGURE 18. Evolution of the streaklines of micro tracers released on the lateral side at $x = 0$, $y = 6r_a/5$ and $8r_a/5$, $z = r_c/8$, together with the 3D iso-surfaces and 2D iso-contours of the concentration of passive scalar. $r_c/R = 3$, $\varepsilon = 40$, $Re = 1$, and $Sc = 100$. (a) 3D representation with iso-surfaces at $\phi = 0.25$, and (b) 2D representation on the central $x$-$z$ plane. Iso-contours are cutoff at $\phi = 0.15$.

On the lateral side, the streaklines exhibit an S-shaped twisted spiral pattern when released from a position closer to the spheroid, as depicted in figure 15. Figure 20 illustrates the evolution of streaklines released at coordinates $x = 0$, $y = 6r_a/5$, $z = r_c/8$. It is observed that the movement directions of the streakline segments released before the spheroid surface reaches the release position and after it passes are opposite in the spheroid's frame of reference. This suggests the presence of a tracer between the two oppositely moving segments that remains stationary relative to the spheroid. These relatively stationary tracers cause the streakline segments released during one tumbling period of the spheroid to form a closed loop with two tracers attached to the spheroid. Such micro tracers attached to the spheroid have also been observed on the central $x$-$z$ plane. As shown in figure 20, a closed loop of streakline segments with four characteristic micro tracers is selected for analysis. Tracers '1' (●) and '3' (●) remain fixed to the spheroid.

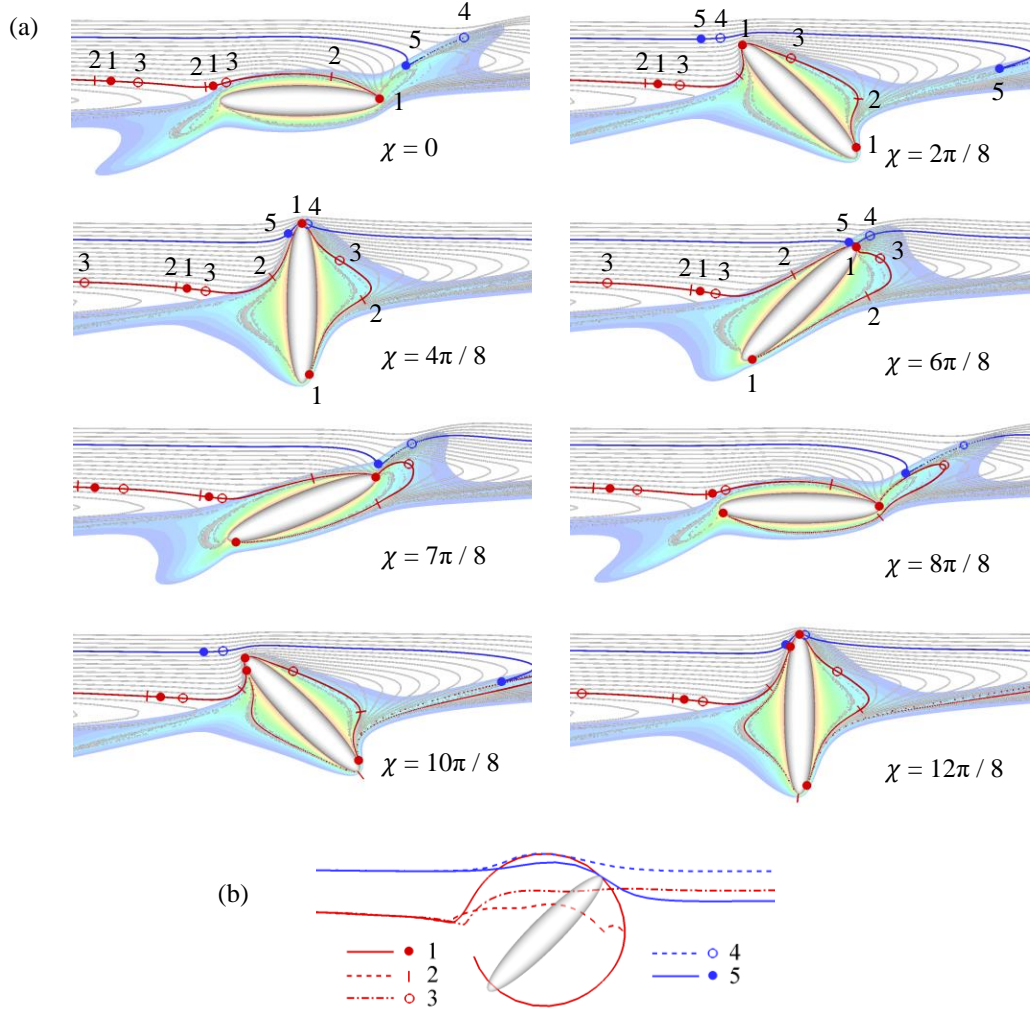

FIGURE 19. Movement of fluid particles represented by micro tracers released at $= -20R$, $y = 0$, $z = 5r_c/20$ and $15r_c/20$ on the central $x$-$z$ plane. $r_c/R = 3$, $\varepsilon = 40$, $Re = 1$, and $Sc = 100$. (a) Streaklines of micro tracers with several characteristic micro tracers marked by numbers, together with the 2D iso-contours of scalar concentration on the central $x$-$z$ plane, and (b) trajectories of the characteristic micro tracers. The 2D iso-contours are cutoff at $\phi = 0.15$.

Influenced by the relative motion of the surrounding fluid, streakline segments '1-2-3' and '3-4-1' move in opposite directions, guided by micro tracers '2' (○) and '4' (○), thus creating a twisted S-shaped spiral pattern. After several cycles, the streakline segments detach from the spheroid and travel downstream along the scalar lines near the central x-z plane. Figure 20(b) shows the trajectories of the four characteristic micro tracers. Tracers '1' and '3' follow the spheroid's tumbling along a fixed ring, while tracers '2' and '4' recirculate along an outward spiral before leaving the spheroid and moving downstream. The non-characteristic tracers follow similar paths to those of '2' and '4' as they depart from the spheroid.

When micro tracers are released slightly further from the spheroid, their behavior changes noticeably. Figure 21 depicts the movement of streakline segments released at coordinates $x = 0$, $y = 8r_a/5$, $z = r_c/8$. At this greater distance, the anisotropic effects of the spheroid's shape in the direction of rotation diminish. Consequently, the more uniform tumbling of the spheroid causes the surrounding fluid to rotate cohesively. This rotation, influenced by centrifugal forces, propels the fluid along an outward spiral, as illustrated in the figure. After several cycles, the fluid particles detach from the spheroid and travel downstream along the scalar lines.

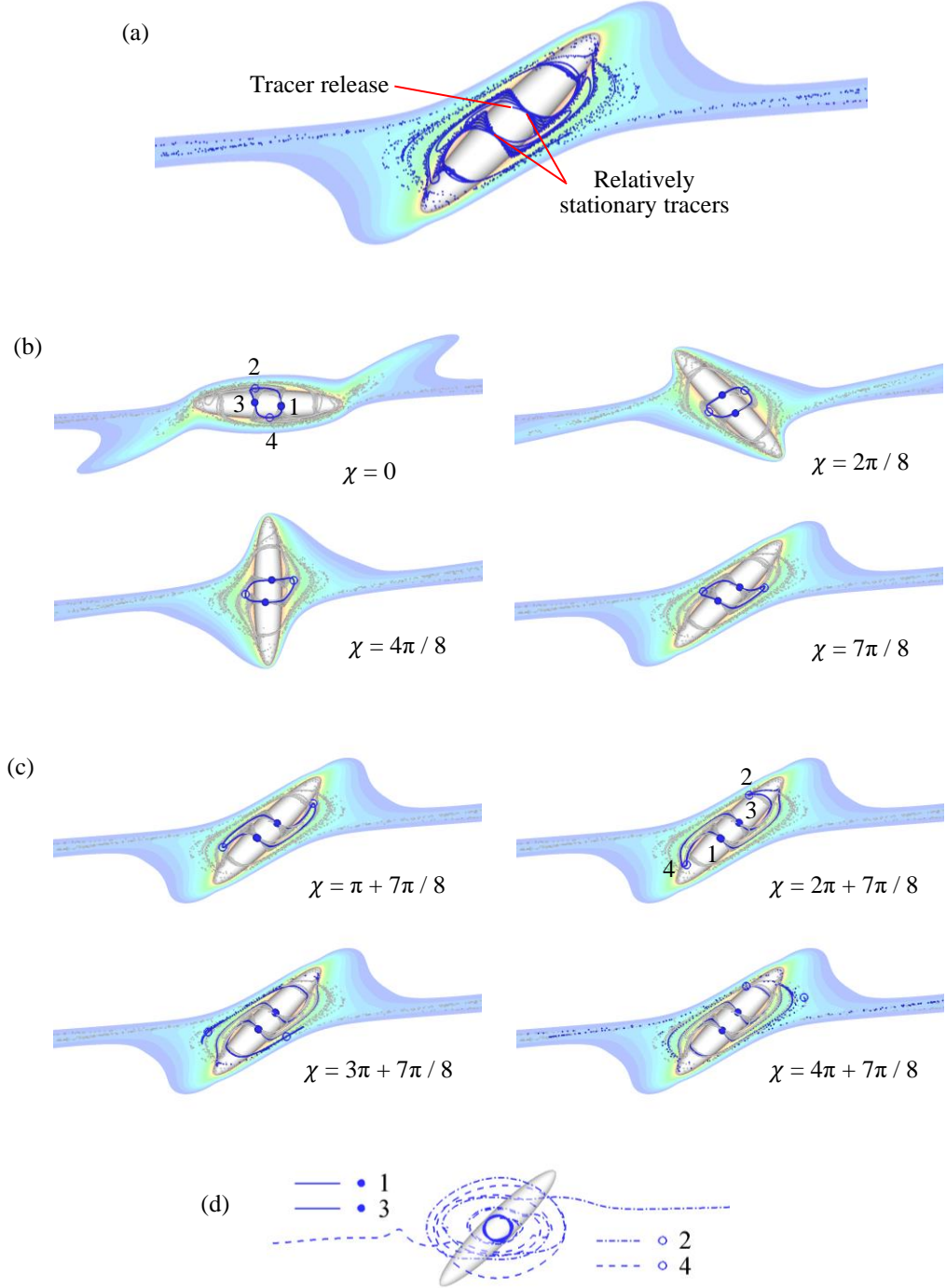

FIGURE 20. Movement of fluid particles represented by micro tracers released on the lateral side at $x = 0$, $y = 6r_a/5$, $z = r_c/8$ of a prolate spheroid with $r_c/R = 3$ and $\varepsilon = 40$. $Re = 1$, and $Sc = 100$. (a) Overall pattern of the streakline of micro tracers, (b) evolution of the streakline over half a rotation period, with several characteristic micro tracers marked by numbers, together with the 2D iso-contours of scalar concentration on the central $x$-$z$ plane, (c) evolution of the streakline over several periods from $\chi = \pi + 7\pi/8$ to $4\pi + 7\pi/8$, and (d) trajectories of the characteristic micro tracers. The 2D iso-contours are cutoff at $\phi = 0.15$.

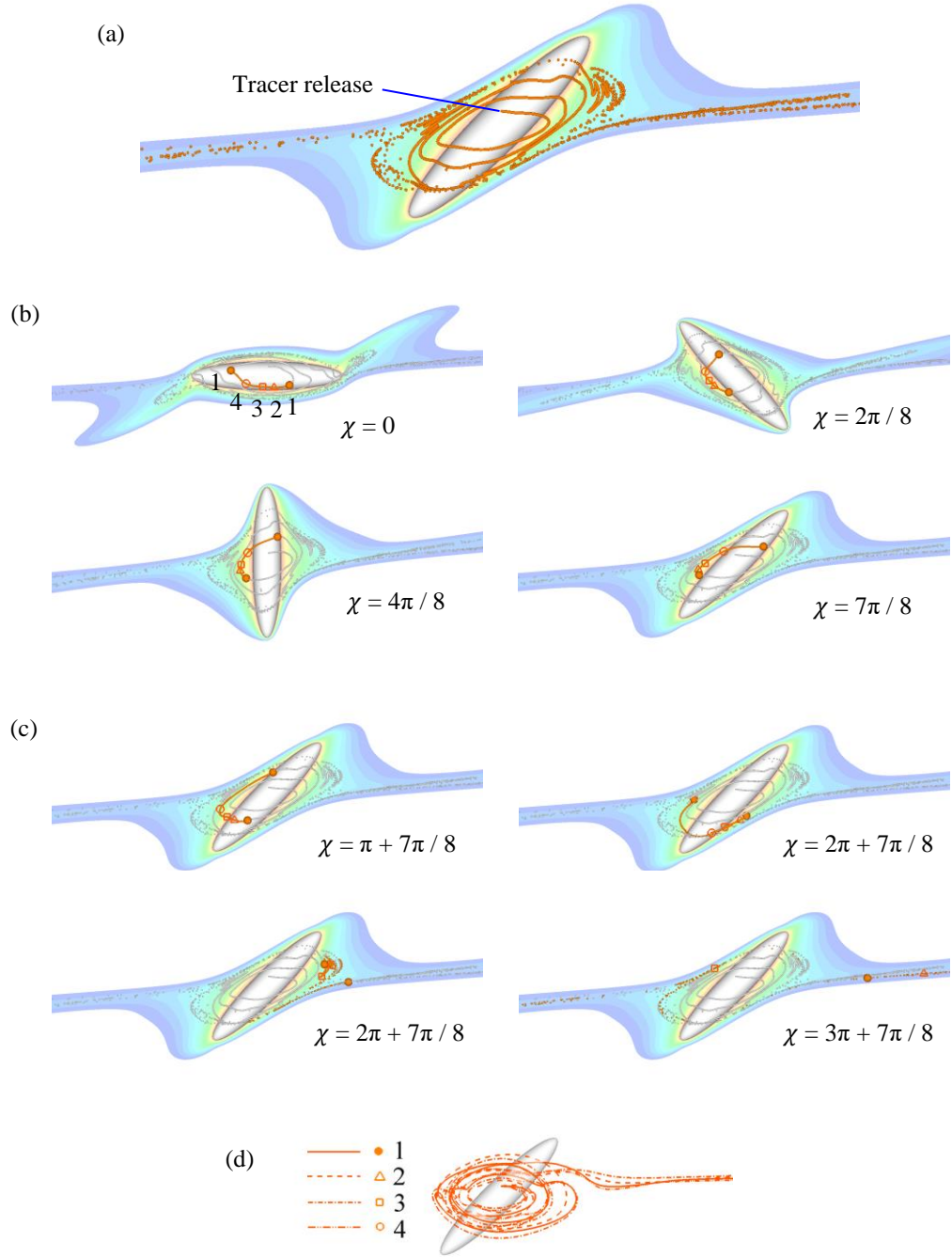

FIGURE 21. Movement of fluid particles represented by micro tracers released on the lateral side at $x = 0$, $y = 8r_a/5$, $z = r_c/8$ of a prolate spheroid with $r_c/R = 3$ and $\varepsilon = 40$. $Re = 1$ and $Sc = 100$. (a) Overall pattern of the streakline of micro tracers, (b) evolution of the streakline over half a rotation period, with several typical points marked by numbers, together with the 2D iso-contours of scalar concentration on the central $x$-$z$ plane, (c) evolution of the streakline over several periods, and (d) trajectories of the typical points. The 2D iso-contours are cutoff at $\phi = 0.15$.

The analysis above underscores the influence of the uniformity of prolate spheroid tumbling on the behavior of the surrounding fluid and the advective transport of passive scalar. Figure 22 illustrates the transition in streakline patterns as the tumbling shifts from relatively non-uniform ($\varepsilon = 4$) to more uniform (($\varepsilon = 40$). This transition encompasses changes in the streakline patterns trapped within the fluid layer and those on the lateral side of the spheroid. For the sake of brevity, detailed descriptions are omitted here.

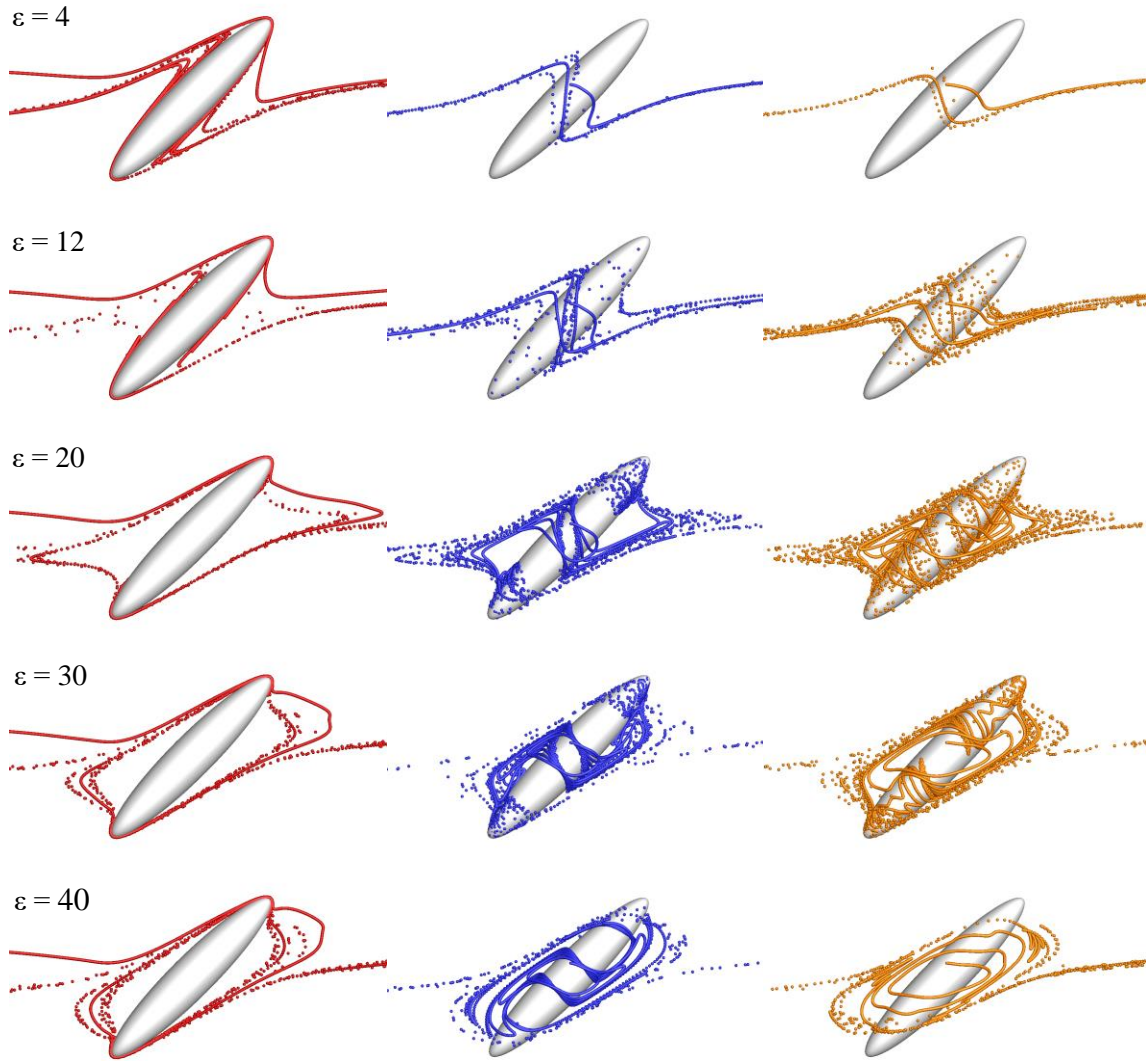

FIGURE 22. Streakline patterns around a prolate spheroid with $r_c/R = 3$ for $Re = 1$. Particle-to-fluid density ratio $\varepsilon$ increases from 4 to 40. Left column: micro tracers released upstream the spheroid on the central x-z plane at $x = -6R, y = 0, z = 5r_c/20$, middle and right: micro tracers released on the lateral side of the spheroid at $x = 0, y = 6r_a/5$ and $8r_a/5, z = r_c/8$.

### 4.3 Quantitative Analysis of Scalar Transport

Variations in the uniformity of prolate spheroid tumbling, influenced by differing particle inertia, lead to distinct flow patterns around the spheroid, thereby impacting the transport rate of passive scalar. In this section, we compare and analyze the scalar transport rates across the specified parameter ranges.

Figure 23 illustrates the variation in the dimensionless scalar transport rate, $Sh$, with respect to dimensionless time $tG$, and rotation angle, $\chi/\pi$, for prolate spheroids characterized by $r_c/R = 3$, and $\varepsilon$ values of 0.4 and 40. The Reynolds number is held constant at $Re = 1$, while the Schmidt number varies from $Sc = 10$ to 100. Following the initial evolution phase, $Sh$ stabilizes into a periodic pattern with respect to both $tG$ and $\chi/\pi$. For both density ratios, an increase in $Sc$ leads to an overall rise in $Sh$. The uniformity of spheroid tumbling significantly influences the temporal patterns of $Sh$, as depicted in figures 23(a) and 23(b). At $\varepsilon = 0.4$, the spheroid lingers near $\chi = n\pi$ for extended periods, resulting in regions where $Sh$ remains constant along the $Sh$ vs. $tG$ curves. Notably, $Sh$ does not reach its minimum at $\chi = n\pi$ but at a slightly smaller angle, as shown in figure 23(c). Conversely, at $\varepsilon = 40$, the spheroid tumbles more uniformly, eliminating these constant $Sh$ regions. The variation of $Sh$ with $\chi/\pi$, presented in figures 23(c) and 23(d), reveals a similar dependency of $Sh$ on $\chi/\pi$ under varying particle inertias. $Sh$ achieves its maximum at an angle slightly less than $(n + 0.5)\pi$ and its minimum at an angle slightly less than $n\pi$. When $\varepsilon = 0.4$, the spheroid's prolonged stay near $\chi = n\pi$ induces a subtle distortion in the curve, as seen in figure 23(c).

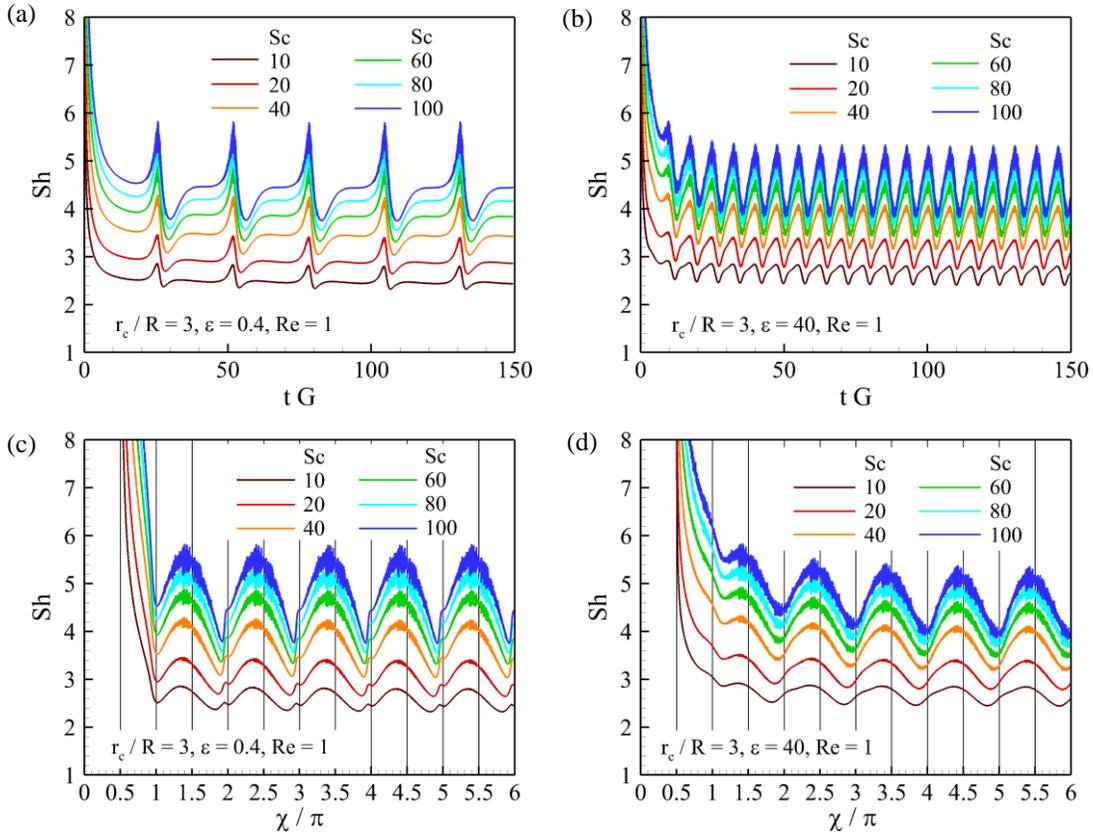

FIGURE 23. Variation of normalized scalar transport rate ($Sh$) with normalized time ($tG$) (a and b) and rotation angle ($\chi/\pi$) (c and d) of a prolate spheroid with $r_c/R = 3$ in a simple shear. for $Re = 1$. (a) $Sh$ vs. $tG$ for $\varepsilon = 0.4$, (b) $Sh$ vs. $tG$ for $\varepsilon = 40$, (c) $Sh$ vs. $\chi/\pi$ for $\varepsilon = 0.4$, and (d) $Sh$ vs. $\chi/\pi$ for $\varepsilon = 40$.

Figure 24 compares the variation in the scalar transport rate, $Sh$, with the rotation angle $\chi/\pi$ for different major semi-axis lengths ($r_c/R$) and particle-to-fluid density ratios ($\varepsilon$). At a Reynolds number ($Re$) of 0.1, fluid inertia is weak, and the particle inertia, characterized by $\varepsilon Re$, is also weak. This results in no significant difference in $Sh$ between $\varepsilon = 0.4$ and $\varepsilon = 40$ for each $r_c/R$, as shown in figure 24(a). An increase in $r_c/R$ leads to a rise in $Sh$, primarily due to the expanded surface area of the spheroid. When $Re$ increases to 1, the enhanced fluid inertia significantly boosts the overall $Sh$, as depicted in figure 24(b). Concurrently, the increased particle inertia causes a noticeable difference in $Sh$ between $\varepsilon = 0.4$ and $\varepsilon = 40$. Except in regions where $\chi$ is near $n\pi$, $Sh$ for $\varepsilon = 0.4$ is higher than for $\varepsilon = 40$. However, this does not imply that the time-averaged $Sh$ is greater for $\varepsilon = 0.4$ than for $\varepsilon = 40$, as the spheroid remains at different angles for varying durations.

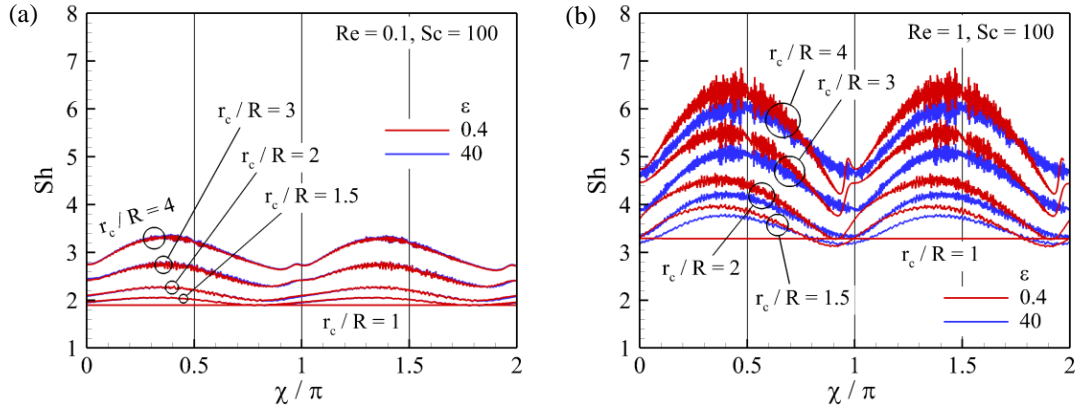

FIGURE 24. Variation of normalized scalar transport rate ($Sh$) with rotation angle ($\chi/\pi$) over half a period of rotation for $Re = 0.1$ and $1$, $\varepsilon = 0.4$ and $40$, and $Sh = 100$. (a) $Re = 1$. (b) $Re = 0.1$.

The overall scalar transport rate can be represented by the time-averaged Sherwood number over one period of spheroid tumbling,

$$\langle Sh \rangle = \frac{1}{T} \int_T Sh \, dt \qquad (4.2)$$

For prolate spheroids, $\langle Sh \rangle$ depends on the combination of $Re$, $r_c/R$, $\varepsilon$, and $Sc$. In many instances, the average scalar transport rate, $\langle Sh \rangle$, is expressed as a function of the Peclet number ($Pe \equiv ReSc$) for a given particle. Figure 25 illustrates the variation of $\langle Sh \rangle$ with $Pe$ across different values of $\varepsilon$, $r_c/R$, and $Re$. The analysis considers two density ratios ($\varepsilon = 0.4$ and $40$) and two Reynolds numbers ($Re = 0.1$ and $1$), with the Schmidt number ($Sc$) ranging from $10$ to $100$. Consequently, $Pe$ spans from $1$ to $10$ for $Re = 0.1$, and from $10$ to $100$ for $Re = 1$. At $Pe = 10$, $\langle Sh \rangle$ for $Re = 1$ and $Sc = 10$ is slightly higher than for $Re = 0.1$ and $Sc = 100$, attributed to the increased fluid inertia. Generally, $\langle Sh \rangle$ rises with $Pe$ for each combination of $\varepsilon$ and $r_c/R$. The results for $r_c/R = 1$ have been thoroughly validated and analyzed in our previous study (Wang & Brasseur 2019). As $r_c/R$ increases, the curve shifts upward while maintaining a similar shape. When $Re = 0.1$, the effect of particle inertia is weak, resulting in no significant difference in $\langle Sh \rangle$ between $\varepsilon = 0.4$ and 40. However, at $Re = 1$, the increased particle inertia significantly raises $\langle Sh \rangle$ for $r_c/R = 3$ and 4 as $\varepsilon$ increases from 0.4 to 40. Figure 24(b) has demonstrated that $Sh$ for $\varepsilon = 0.4$ is higher than for $\varepsilon = 40$ at most rotation angles, except near $\chi = n\pi$. Nonetheless, the spheroid remains near $\chi = n\pi$ longer when $\varepsilon = 0.4$ than when $\varepsilon = 40$, leading to a lower $\langle Sh \rangle$ for $\varepsilon = 0.4$ compared to $\varepsilon = 40$.

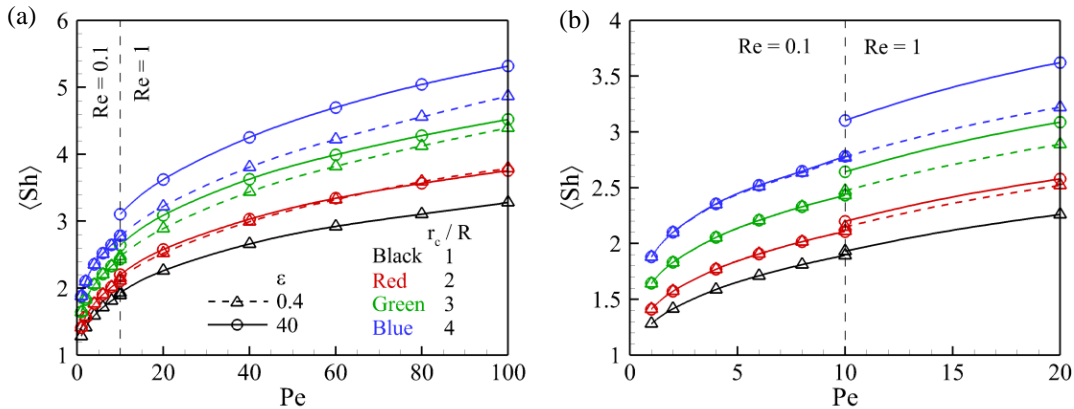

FIGURE 25. Variation of time-averaged scalar transport rate ($\langle Sh \rangle$) with Peclet number ($Pe$) for various length of major semi-axis of prolate spheroids. (a) Full range of $Pe$ considered in this study, (b) range of smaller $Pe$. The symbols represent the results of LBM simulation.

Compared to the complexities introduced by fluid and particle inertia, the influence of particle shape is relatively straightforward. Wang et al. (2022) developed a theoretical solution for diffusion-driven mass transport from spheroidal particles, both prolate and oblate, and systematically examined the effect of particle shape. By normalizing the time-averaged scalar transport rate $\langle Sh \rangle$ from this study with the diffusion-driven transport rate $Sh_{dif}$ from Wang et al. (2022), the impact of particle shape on $\langle Sh \rangle$ can be minimized. Figure 26 presents the normalized scalar transport rate $\langle Sh \rangle/Sh_{dif}$ as a function of the Peclet number $Pe$ for various $r_c/R$ and $\varepsilon$ values. At $Re = 0.1$, neither fluid nor particle inertia significantly affects scalar transport when particle shape changes, resulting in a nearly identical dependence of $\langle Sh \rangle/Sh_{dif}$ on $Pe$ across different $r_c/R$ values. The difference in $\langle Sh \rangle/Sh_{dif}$ between $r_c/R = 4$ and 1 is less than 3%. However, at $Re = 1$, the combined effects of fluid and particle inertia lead to a noticeable increase in $\langle Sh \rangle/Sh_{dif}$ as $r_c/R$ increases from 1 to 4. These findings suggest that when both fluid and particle inertia are weak, normalizing the scalar transport rate with the corresponding diffusion-driven transport rate can mitigate the effect of particle shape to some extent. Consequently, at low Reynolds numbers, the scalar transport rate of a prolate spheroid can be predicted from the results for a sphere using a consistent relationship between $\langle Sh \rangle/Sh_{dif}$ and $Pe$. This aspect is not the primary focus of this work, so it is not explored further.

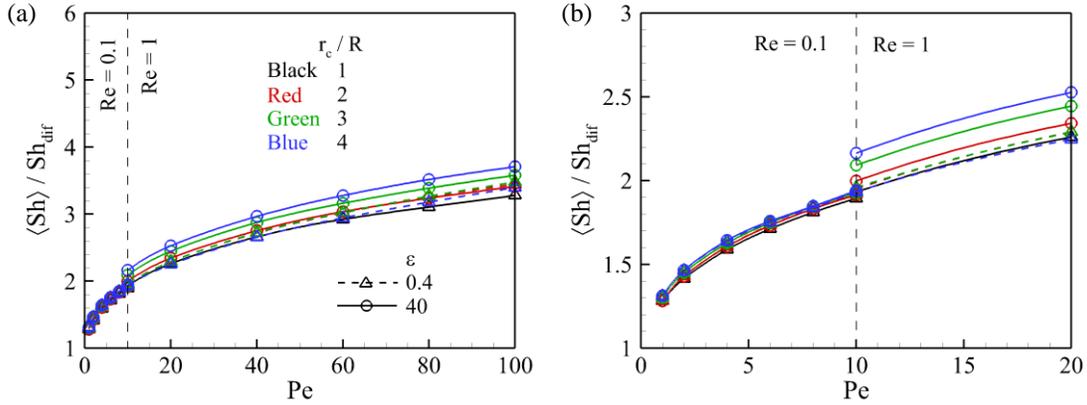

FIGURE 26. Time-averaged scalar transport rate normalized by diffusive scalar transport rate ($\langle Sh \rangle/Sh_{dif}$) versus Peclet number ($Pe$) for various length of major semi-axis of prolate spheroid. (a) Full range of $Pe$ considered in this study, (b) range of smaller $Pe$.

## 5. Conclusions

Using high-fidelity numerical simulations based on a lattice Boltzmann framework, an in-depth study has been conducted on the heat and mass transport from a prolate spheroid suspended in a simple shear flow. In these simulations, temperature and mass concentration are treated as passive scalars released from the spheroid's surface. This research primarily focuses on identifying the flow modes induced by the tumbling of the spheroid at lower Reynolds numbers and the resulting patterns of scalar transport.

An extensive analysis of the tumbling angular velocity of prolate spheroids with varying aspect ratios and particle-to-fluid density ratios highlights the crucial influence of particle inertia on the uniformity of spheroid tumbling. This uniformity, in turn, impacts the flow patterns surrounding the spheroids. When particle inertia is weak, the angular velocity varies in a non-uniform manner. Conversely, strong particle inertia results in a more uniform angular velocity. Considering the aspect ratio of the spheroid, the tumbling behavior can be classified into three categories based on the uniformity of angular velocity: relatively uniform tumbling in nearly spherical prolate spheroids, relatively non-uniform tumbling in slender spheroids with weak particle inertia, and relatively uniform tumbling in slender prolate spheroids with strong particle inertia. Each category generates distinct flow modes around the spheroid.

For both relatively uniform and non-uniform tumbling of slender spheroids, the tumbling motion generates a scalar line in the fluid with a higher concentration of passive scalar. These newly formed scalar lines, driven by the spheroid's poles, sweep through the wake regions and merge with clusters of previously generated scalar lines. As fluid flows past the spheroid, it transports the passive scalar downstream along these lines, illustrating the primary pathway for passive scalar transport via flow advection. Near the central streamwise plane, some fluid flows directly downstream along the scalar lines, while other portions become trapped in the fluid layer on the spheroid's surface. This trapped fluid undergoes a repetitive process where part of it continues downstream and part becomes trapped again. The variation in tumbling uniformity, influenced by particle inertia, results in different fluid movement patterns downstream or within the fluid layer. On the lateral sides, during relatively non-uniform tumbling, most fluid travels directly downstream along the scalar lines. In contrast, during relatively uniform tumbling, the fluid rotates through several cycles in a twisted or normal outward spiral before proceeding downstream.

The transport rate of passive scalar from a prolate spheroid results from the interplay of various flow modes around the spheroid, each influenced by factors such as spheroid aspect ratio, fluid inertia, and particle inertia, all interacting in a complex manner. In equilibrium, the scalar transport rate exhibits periodic variations with time and rotation angle. For prolate spheroids with the same aspect ratio, the scalar transport rate is generally higher when particle inertia is weak, except at rotation angles where the spheroid's major axis aligns with the flow direction. However, because the spheroid spends more time at these angles when particle inertia is weak, the time-averaged scalar transport rate is lower compared to when particle inertia is strong. Overall, increases in fluid inertia, particle inertia, and spheroid aspect ratio lead to a rise in the time-averaged scalar transport rate within the parameter ranges considered in this study. Additionally, when both fluid and particle inertia are weak, the scalar transport rates for different aspect ratios, when normalized by the corresponding diffusion-driven transport rate, converge to a similar dependence on the Peclet number. This finding aids in the development of future empirical models for heat and mass transport in particle systems.

This paper aims to lay the groundwork for future research on heat and mass transport resulting from more complex particle motions across a wider range of parameters.

**Acknowledgements:** This work was supported by the National Science Foundation (Award ID: 2138740).

**Declaration of Interests.** The authors report no conflict of interest.